\documentclass[11pt,a4paper,aps,pra]{revtex4-1}
\def\qe{{\sc Quantum ESPRESSO}}

\def\exec#1{\texttt{#1}}

\def\var#1{\texttt{#1}}
\DeclareMathAlphabet\mathbfcal{OMS}{cmsy}{b}{n}

\def\PWscf{\normalfont\texttt{PWscf}}
\def\atomic{\normalfont\texttt{atomic}}
\def\PWneb{\normalfont\texttt{PWneb}}
\def\PHonon{\normalfont\texttt{PHonon}}
\def\PostProc{\normalfont\texttt{PostProc}}
\def\XSPECTRA{\normalfont\normalfont\texttt{XSpectra}}
\def\pwtk{\normalfont\texttt{pwtk}}
\def\Pwtk{\normalfont\texttt{Pwtk}}
\def\CP{\normalfont\texttt{CP}}
\def\FD{\normalfont\texttt{FD}}
\def\Plumed{\normalfont\texttt{Plumed}}
\def\EPW{\normalfont\texttt{EPW}}
\def\GWL{\normalfont\texttt{GWL}}
\def\SternheimerGW{\normalfont\texttt{SternheimerGW}}
\def\Environ{\normalfont\texttt{Environ}}
\def\turboTDDFT{\normalfont\texttt{turboTDDFT}}
\def\turboEELS{\normalfont\texttt{turboEELS}}
\def\QE-GIPAW{\normalfont\texttt{QE-GIPAW}}
\def\thermopw{\normalfont\texttt{thermo\_pw}}
\def\thermal2{\normalfont\texttt{thermal2}}
\def\d3q{\normalfont\texttt{d3q}}
\def\yambo{\normalfont\texttt{yambo}}
\def\sax{\normalfont\texttt{SaX}}
\def\BerkeleyGW{\normalfont\texttt{BerkeleyGW}}
\def\WanT{\normalfont\texttt{WanT}}
\def\Wannier90{\normalfont\texttt{Wannier90}}
\def\AiiDA{\normalfont\texttt{AiiDA}}

\def\QEmodes{\texttt{QE-emacs-modes}}

\def\pwmode{\texttt{pw-mode}}

\def\rr{{\bf r}}
\def\kk{{\bf k}}
\def\qq{{\bf q}}
\def\GG{{\bf G}}
\def\RR{\mathbf{R}}
\def\dr{d\rr}
\def\dk{d\kk}
\def\ie{\emph{i.e.}}
\def\eg{\emph{e.g.}}

\usepackage{graphicx}
\usepackage{color, colortbl}
\usepackage{multirow}
\usepackage[table]{xcolor}
\usepackage{amsmath}
\usepackage[normalem]{ulem}
\usepackage{natbib}

\usepackage{tikz}
\pgfrealjobname{QEfinal}

\begin{document} 

\nocite{*}
\author{P. Giannozzi$^a$, O. Andreussi$^{b,i}$, T. Brumme$^c$,
  O. Bunau$^d$, M. Buongiorno Nardelli$^e$, M. Calandra$^d$, R. Car$^f$,
  C. Cavazzoni$^g$, D. Ceresoli$^h$, M. Cococcioni$^i$, N. Colonna$^i$,
  I. Carnimeo$^a$, A. Dal Corso$^j$, S. de Gironcoli$^j$, P. Delugas$^j$, 
  R. A. DiStasio Jr.$^k$, A. Ferretti$^l$, A. Floris$^m$, G. Fratesi$^n$,
  G. Fugallo$^o$, R. Gebauer$^p$, U. Gerstmann$^q$, F. Giustino$^r$,
  T. Gorni$^j$, J. Jia$^k$, M. Kawamura$^s$, H.-Y. Ko$^f$,
  A. Kokalj$^t$, E. K\"u\c{c}\"ukbenli$^j$, M. Lazzeri$^d$,
  M. Marsili$^u$, N. Marzari$^i$,
  F. Mauri$^v$, N. L. Nguyen$^j$, H.-V. Nguyen$^w$, A. Otero-de-la-Roza$^x$,
  L. Paulatto$^d$, S. Ponc\'e$^r$, D. Rocca$^{y,z}$, R. Sabatini$^1$,
  B. Santra$^f$, M. Schlipf$^r$, A. P. Seitsonen$^{2,3}$, A. Smogunov$^4$,
  I. Timrov$^i$, T. Thonhauser$^5$, P. Umari$^{u,6}$, N. Vast$^7$,
  X. Wu$^8$, S. Baroni $^j$
}

\affiliation{(a) Dept.\ of Mathematical, Physical, and Computer Sciences,
  University of Udine, via delle Scienze 206, I-33100 Udine, Italy\\
  (b) Institute of Computational Sciences, Universit\`a della Svizzera
  Italiana, Lugano, Svizzera\\
  (c) Wilhelm-Ostwald-Institute of Physical and Theoretical Chemistry, Leipzig University, Linn\'estr. 2, D-04103 Leipzig, Germany\\
  (d) IMPMC, UMR CNRS 7590, Sorbonne Universit\'es-UPMC University
  Paris 06, MNHN, IRD, 4 Place Jussieu, F-75005 Paris, France\\
  (e) Department of Physics and Department of Chemistry, University of North Texas, Denton, USA\\
  (f) Department of Chemistry, Princeton University, Princeton, NJ 08544, USA\\
  (g) CINECA - Via Magnanelli 6/3, I-40033 Casalecchio di Reno, Bologna, Italy\\
  (h) Institute of Molecular Science and Technologies (ISTM),
  National Research Council (CNR), I-20133 Milano, Italy\\
  (i) Theory and Simulation of Materials (THEOS), and National Centre for Computational Design and Discovery of Novel Materials (MARVEL), Ecole Polytechnique F\'ed\'erale de Lausanne,
  CH-1015 Lausanne, Switzerland\\
  (j) SISSA-ISAS, via Bonomea, 265, I-34136 Trieste, Italy\\
  (k) Department of Chemistry and Chemical Biology, Cornell University, Ithaca, NY 14853 USA\\
  (l) CNR Istituto Nanoscienze, I-42125 Modena, Italy\\
  (m) University of Lincoln, UK
  (n) Dipartimento di Fisica, Universit\`a degli Studi di Milano, via Celoria 16, I-20133 Milano, Italy
  (o) ETSF,  Laboratoire des Solides Irradi\'es, Ecole Polytechnique, F-91128 Palaiseau cedex, France\\
  (p) The Abdus Salam International Centre for Theoretical Physics (ICTP), Strada Costiera 11, I-34151 Trieste, Italy\\
  (q) Department Physik, Universit\"at Paderborn, D-33098 Paderborn, Germany\\
  (r) Department of Materials, University of Oxford, Parks Road, Oxford OX1 3PH, United Kingdom\\
  (s) The Institute for Solid State Physics, Kashiwa, Japan\\
  (t) Department of Physical and Organic Chemistry, Jo\v{z}ef Stefan Institute, Jamova 39, 1000 Ljubljana, Slovenia\\
  (u) Dipartimento di Fisica e Astronomia, Universit\`a di Padova,  via Marzolo 8, I-35131 Padova, Italy\\
  (v) Dipartimento di Fisica, Universit\`a di Roma La Sapienza, Piazzale Aldo Moro 5, I-00185 Roma, Italy\\
  (w) Institute of Physics, Vietnam Academy of Science and Technology, 10 Dao Tan, Hanoi, Vietnam\\
  (x) Department of Chemistry, University of British Columbia, Okanagan, Kelowna BC V1V 1V7, Canada\\
  (y) Universit\'e de Lorraine, CRM\textsuperscript{2}, UMR 7036, F-54506 Vandoeuvre-l\`es-Nancy, France\\
  (z) CNRS, CRM\textsuperscript{2}, UMR 7036, F-54506 Vandoeuvre-l\`es-Nancy,   France\\
  (1) Orionis Biosciences, Boston
  (2) Institut f\"ur Chimie, Universit\"at Zurich, CH-8057 Z\"urich,
      Switzerland\\
  (3) D\'epartement de Chimie, \'Ecole Normale Sup\'erieure, F-75005 Paris, France\\
  (4) SPEC, CEA, CNRS, Universit\'e Paris-Saclay, F-91191 Gif-Sur-Yvette, France\\
  (5) Department of Physics, Wake Forest University, Winston-Salem, NC 27109, USA\\
  (6) CNR-IOM DEMOCRITOS, Istituto Officina dei Materiali, Consiglio Nazionale delle Ricerche\\
  (7) Laboratoire des Solides Irradi\'es, \'Ecole Polytechnique, CEA-DRF-IRAMIS, CNRS UMR 7642, Universit\'e Paris-Saclay, F-91120 Palaiseau, France\\
  (8) Department of Physics, Temple University, Philadelphia, PA 19122-1801, USA
   }

\date{\today}
\title{Advanced capabilities for materials modelling with \qe}
\begin{abstract}
  \qe\ is an integrated suite of open-source computer codes for
  quantum simulations of materials using state-of-the art
  electronic-structure techniques, based on density-functional theory,
  density-functional perturbation theory, and many-body perturbation
  theory, within the plane-wave  pseudo-potential and
  projector-augmented-wave approaches. \qe\ owes its popularity to the
  wide variety of properties and processes it allows to simulate, to
  its performance on an increasingly broad array of hardware
  architectures, and to a community of researchers that rely on its
  capabilities as a core open-source development platform to implement
  theirs ideas. In this paper we describe recent extensions and
  improvements, covering new methodologies and property calculators,
  improved parallelization, code modularization, and extended
  interoperability both within the distribution and with external
  software.  
\end{abstract}

\maketitle
\tableofcontents

\section{Introduction}

Numerical simulations based on density-functional theory (DFT)
\cite{Hohenberg1964,Kohn1965} have become a powerful and widely used
tool for the study of materials properties. Many of such simulations
are based upon the ``plane-wave pseudopotential method'', often using
ultrasoft pseudopotentials \cite{Vanderbilt1990} or the projector
augmented wave method (PAW) \cite{Bloechl1994} (in the following, all
of these modern developments will be referred to under the generic
name of ``pseudopotentials''). An important role in the diffusion of
DFT-based techniques has been played by the availability of robust and
efficient software implementations \cite{Lejaeghereaad3000}, as is the
case for {\sc Quantum ESPRESSO}, which is an open-source software
distribution---\ie, an integrated suite of codes---for
electronic-structure calculations based on DFT or many-body
perturbation theory, and using plane-wave basis sets and
pseudopotentials \cite{QE:2009}.

  The core philosophy of \qe\ can be summarized in four keywords:
  openness, modularity, efficiency, and innovation. The distribution
  is based on two core packages, \PWscf\ and \CP, performing self-consistent
  and molecular-dynamics calculations respectively, and on additional packages
  for more advanced calculations. Among these we quote in particular:
  \PHonon, for linear-response calculations of vibrational properties;
  \PostProc, for data analysis and postprocessing;
  \atomic, for pseudopotential generation;
  \XSPECTRA, for the calculation of X-ray absorption spectra;
  \texttt{GIPAW}, for nuclear magnetic resonance and electron
  paramagnetic resonance calculations. 

In this paper we describe and document the novel or improved
capabilities of \qe\ up to and including version 6.2. We do
not cover features already present in v.4.1 and described in
Ref. \onlinecite{QE:2009}, to which we refer for further details. The
list of enhancements includes theoretical and methodological
extensions but also performance enhancements for current parallel
machines and modularization and extended interoperability with other
software. 

Among the theoretical and methodological extensions, we mention in particular:
\begin{itemize}
\item Fast implementations of exact (Fock) exchange for hybrid
  functionals \cite{Lin:2016,Wu_2009,DiStasio_2014,Ko_2017};
  implementation of non-local
  van der Waals functionals \cite{Berland_2015:van_waals} and of
  explicit corrections for van der Waals interactions
  \cite{grimme2006,DFT-TS,johnson2007,ncibook}; improvement and
  extensions of Hubbard-corrected functionals
  \cite{sclauzero13,himmetoglu11}. 
\item Excited-state calculations within time-dependent
  density-functional and many-body perturbation theories. 
\item Relativistic extension of the PAW formalism, including
  spin-orbit interactions in density-functional
  theory\cite{Dalcorso:2010,Dalcorso:2012}.  
\item Continuum embedding environments (dielectric solvation models,
  electronic enthalpy, electronic surface tension, periodic boundary
  corrections) via the \Environ\ module
  \cite{Andreussi:2012,Andreussi:2014} and its time-dependent
    generalization \cite{Timrov:2015b}.
\end{itemize}
Several new packages, implementing the calculation of new properties,
have been added to \qe. We quote in particular:
\begin{itemize}
\item \turboTDDFT\ \cite{Walker:2006, Rocca:2008,
  Malcoglu:2011,Ge:2014} and \turboEELS\ \cite{Timrov:2013,
  Timrov:2015}, for excited-state calculations within time-dependent
  DFT (TDDFT), without computing virtual orbitals, also interfaced
    with the \Environ\ module (see above).
\item \QE-GIPAW, replacing the old \texttt{GIPAW} package, for nuclear
  magnetic resonance and electron paramagnetic resonance
  calculations. 
\item \EPW, for electron-phonon calculations using Wannier-function
  interpolation \cite{Ponce:2016}. 
\item \GWL\ and \SternheimerGW\ for quasi-particle and excited-state
  calculations within many-body perturbation theory, without computing
  any virtual orbitals, using the Lanczos bi-orthogonalization
  \cite{Umari:2009,Umari:2010} and multi-shift conjugate-gradient
  methods \cite{sternheimergw}, respectively. 
\item \thermopw, for computing thermodynamical properties in the
  quasi-harmonic approximation, also featuring an advanced
  master-slave distributed computing scheme, applicable to generic
  high-throughput calculations \cite{thermopw}. 
\item \d3q\ and \thermal2, for the calculation of anharmonic 3-body
  interatomic force constants, phonon-phonon interaction and thermal
  transport \cite{paulattoPRB,fugalloPRB}. 
\end{itemize}
Improved parallelization is crucial to enhance performance and to
fully exploit the power of modern parallel architectures. A careful
removal of memory bottlenecks and of scalar sections
of code is a pre-requisite for better and extending scaling. 
Significant improvements have been achieved, in particular for
hybrid functionals \cite{Varini:2013,Nersc:2017}.

Complementary to this, a complete pseudopotential library,
\texttt{pslibrary}, including fully-relativistic pseudopotentials,
has been generated \cite{pslibrary,Dalcorso:2015}. A 
curation effort \cite{sssp} on all the pseudopotential libraries 
available for \qe\ has led to the identification of optimal
pseudopotentials for efficiency or for accuracy in the calculations,
the latter delivering an agreement comparable to any of the best
all-electron codes \cite{Lejaeghereaad3000}. Finally, a significant
effort has been dedicated to modularization and to enhanced
interoperability with other software. The structure of the
distribution has been revised, the code base has been re-organized,
the format of data files re-designed in line with modern standards. As
notable examples of interoperability with other software, we
mention in particular the interfaces with the \texttt{LAMMPS}
molecular dynamics (MD) code \cite{Plimpton:1995} used as
molecular-mechanics ``engine'' in the \qe\ implementation of the QM-MM
methodology \cite{Ma:2015}, and with the \texttt{i-PI} MD driver
\cite{Ceriotti:2014}, also featuring path-integral MD.  

All advances and extensions that have not been documented elsewhere are described in the next sections. For more details on new packages we refer to the respective references.

The paper is organized as follows. Sec. II contains a description of new theoretical and methodological developments and of new packages distributed together with \qe. Sec. III contains a description of improvements of parallelization, updated information on the philosophy and general organization of \qe, notably in the field of modularization and interoperability. Sec. IV contains an outlook of future directions and our conclusions.

\section{Theoretical, algorithmic, and methodological extensions}

In the following, CGS units are used, unless noted otherwise.

\subsection{Advanced functionals}

\subsubsection{Advanced implementation of exact (Fock) exchange and hybrid functionals}
\label{sec:exx}

Hybrid functionals are already the \emph{de facto} standard in quantum
chemistry and are quickly gaining popularity in the condensed-matter
physics and computational materials science communities. Hybrid
functionals reduce the self-interaction error that plagues lower-rung
exchange-correlation functionals, thus achieving more accurate and 
reliable predictive capabilities. This is 
of particular importance in the calculation of orbital energies, which
are an essential ingredient in the treatment of band alignment and
charge transfer in heterogeneous systems, as well as the input
for higher-level electronic-structure calculations based on many-body
perturbation theory. However, the widespread use of hybrid functionals
is hampered by the often prohibitive computational requirements of the
exact-exchange (Fock) contribution, especially when working with
a plane-wave basis set. The basic ingredient here is the action $(\hat
V_x\phi_i)(\rr)$ of the Fock operator $\hat V_x$ onto a
(single-particle) electronic state $\phi_i$, requiring a sum over all
occupied Kohn-Sham (KS) states $\{\psi_j\}$. For spin-unpolarized
systems, one has: 
\begin{equation}
  \label{eq:Vx}
  (\hat V_x\phi_i)(\rr) = -e^2\sum_j\psi_j(\rr)\int \dr'
        \frac{\psi^*_j(\rr')\phi_i(\rr')} {|\rr-\rr'|},
\end{equation}
where $-e$ is the charge of the electron. In the original
algorithm~\cite{QE:2009} implemented in \PWscf, self-consistency is
achieved \textit{via} a double loop: 
in the inner one the $\psi$'s entering the definition of the Fock
operator in Eq. \eqref{eq:Vx} are kept fixed, while the outer one
cycles until the Fock operator converges to within a given threshold. 
In the inner loop, the integrals appearing in Eq.~(\ref{eq:Vx}): 
\begin{equation}
  \label{eq:Poisson}
  v_{ij}(\rr) = \int \dr' \frac{\rho_{ij}(\rr')}{|\rr-\rr'|},\qquad
  \rho_{ij}(\rr) = \psi^*_i(\rr)\phi_j(\rr),
\end{equation}
are computed by solving the Poisson equation in reciprocal space using
fast Fourier transforms (FFT). This algorithm is straightforward but
slow, requiring ${\cal O}\bigl ((N_bN_k)^2\bigr )$ FFTs, where $N_b$
is the number of electronic states (``bands'' in solid-state parlance)
and $N_k$ the number of $\kk$ points in the Brillouin zone (BZ). While
feasible in relatively small cells, this unfavorable scaling with the
system size makes calculations with hybrid functionals challenging if
the unit cell contains more than a few dozen atoms.  

To enable exact-exchange calculations in the condensed phase, various
ideas have been conceived and implemented in recent
\qe\ versions. Code improvements aimed at either optimizing or better 
parallelizing the standard algorithm are described in
Sec.~\ref{sec:newpara}. In this section we describe two important
algorithmic developments in \qe, both entailing a significant
reduction in the computational effort: the \emph{adaptively
compressed exchange} (ACE) concept~\cite{Lin:2016} and a
linear-scaling (${\cal O}(N_b)$) framework for performing
hybrid-functional \textit{ab initio} molecular dynamics using
maximally localized Wannier functions
(MLWF)~\cite{Wu_2009,DiStasio_2014,Ko_2017}.

\paragraph{Adaptively compressed exchange {\label{ACE}}}

The simple formal derivation of ACE allows for a robust implementation, which applies straightforwardly both to isolated or aperiodic systems ($\Gamma-$only sampling of the BZ, that is, $\kk=0$) and to periodic ones (requiring sums over a grid of $\kk$ points in the BZ); to norm conserving and ultrasoft pseudopotentials or PAW; to spin-unpolarized or polarized cases or to non-collinear magnetization.  Furthermore, ACE is compatible with, and takes advantage of, all available parallelization levels implemented in \qe: over plane waves, over $\kk$ points, and over bands.

With ACE, the action of the exchange operator is rewritten as 
\begin{equation}
  \label{eq:Vxace}
  |\hat V_x\phi_i\rangle \simeq \sum_{jm}|\xi_j\rangle
  (M^{-1})_{jm} \langle \xi_m|\phi_i\rangle,
\end{equation}
where $|\xi_i\rangle = \hat V_x |\psi_i\rangle $ and
  $M_{jm}=\langle \psi_j|\xi_m\rangle $.
At self-consistency, ACE becomes exact for $\phi_i$'s in the occupied
manifold of KS states. 
It is straightforward to implement ACE in the double-loop
structure of \PWscf. The new algorithm is significantly faster while
not introducing any loss of accuracy at convergence. Benchmark tests
on a single processor show a $3\times$ to $4\times$ speedup for typical
calculations in molecules, up to $6\times$ in extended systems~\cite{Ivan:2017}.

An additional speedup may be achieved by using a reduced FFT cutoff in
the solution of Poisson equations.  In Eq.~(\ref{eq:Vx}), the exact
FFT algorithm requires a FFT grid containing G-vectors up to a modulus
$G_{max}=2G_c$, where $G_c$ is the largest modulus of G-vectors in the
plane-wave basis used to expand $\psi_i$ and $\phi_j$, or, in terms of
kinetic energy cutoff, up to a cutoff $E_x=4E_c$, where $E_c$ is the
plane-wave cutoff. The presence of a $1/G^2$ factor in the reciprocal
space expression suggests, and experience confirms, that this
condition can be relaxed to $E_x\sim 2E_c$ with little loss of
precision, down to $E_x= E_c$ at the price of increasing somewhat this
loss \cite{Marsili:2013}. The kinetic-energy cutoff for Fock-exchange computations can be
tuned by specifying the keyword \var{ecutfock} in input. 
  
Hybrid functionals have also been extended to the case of ultrasoft
pseudopotentials and to PAW, following the method of
Ref.~\onlinecite{Paier:2005}. A large number of integrals involving
augmentation charges $q_{lm}$ are needed in this case, thus offsetting
the advantage of a smaller plane-wave basis set. Better performances
are obtained by exploiting the localization of the $q_{lm}$ and
computing the related terms in real space, at the price of small
aliasing errors. 

These improvements allow to significantly speed up a calculation, or
to execute it on a larger number of processors, thus extending the
reach of calculations with hybrid functionals. The bottleneck
represented by the sum over bands and by the FFT in Eq.~(\ref{eq:Vx})
is however still present: ACE just reduces the number of such
expensive calculations, but doesn't eliminate them. In order to
achieve a real breakthrough, one has to get rid of delocalized bands
and FFT's, moving to a representation of the electronic structure
in terms of localized orbitals. Work along  this line using the
\emph{selected column density matrix} localization scheme
\cite{LinSCDM:2015,LinSCDM:2016} is ongoing. In the next section we
describe a different approach, implemented in the \CP\ code, based on
maximally localized Wannier functions (MLWF). 

\paragraph{Ab-initio molecular dynamics using maximally localized Wannier functions \label{EXX-MLWF}}

The \CP\ code can now perform highly efficient hybrid-functional
\textit{ab initio} MD using MLWFs \cite{PhysRevB.56.12847}
$\{\overline{\varphi}_{i}\}$ to represent the occupied space, instead
of the canonical KS orbitals $\{\psi_{i}\}$, which are typically
delocalized over the entire simulation cell. The MLWF localization
procedure can be written as a unitary transformation,
$\overline{\varphi}_{i}(\rr) = \sum_{j} U_{ij} \psi_{j}(\rr)$, where
$U_{ij}$ is computed at each MD time step by minimizing the total
spread of the orbitals via a second-order damped dynamics scheme,
starting with the converged $U_{ij}$ from the previous time step as
initial guesses~\cite{sharma_2003}.  
  
The natural sparsity of the exchange interaction provided by a
localized representation of the occupied orbitals (at least in systems
with a finite band gap) is efficiently exploited during the evaluation
of exact-exchange based applications (\eg, hybrid DFT
functionals). This is accomplished by computing each of the required
pair-exchange potentials $\overline{v}_{ij}(\rr)$ (corresponding to a
given localized pair-density $\overline{\rho}_{ij}(\rr)$) through
the numerical solution of the Poisson equation:
\begin{equation}
\nabla^{2} \overline{v}_{ij}(\rr) = -4\pi \overline{\rho}_{ij}(\rr),
\qquad \overline{\rho}_{ij}(\rr) = \overline{\varphi}_{i}^{\ast}(\rr)
\overline{\varphi}_{j}(\rr) 
\end{equation}
using finite differences on the real-space grid. Discretizing the Laplacian operator ($\nabla^2$) using a 19-point central-difference stencil (with an associated $\mathcal{O}(h^6)$ accuracy in the grid spacing $h$), the resulting sparse linear system of equations is solved using the conjugate-gradient technique subject to the boundary conditions imposed by a multipolar expansion of
$\overline{v}_{ij}(\rr)$: 
\begin{equation}
\overline{v}_{ij}(\rr) = 4\pi \sum_{lm} \frac{Q_{lm}}{2l+1}
\frac{Y_{lm}(\theta, \phi)}{r^{l+1}}, \qquad Q_{lm} = \int \dr
Y_{lm}^{\ast}(\theta, \phi) r^{l} \overline{\rho}_{ij}(\rr) 
\end{equation}
in which the $Q_{lm}$ are the multipoles describing
$\overline{\rho}_{ij}(\rr)$~\cite{Wu_2009,DiStasio_2014,Ko_2017}. 

Since $\overline{v}_{ij}(\rr)$ only needs to be evaluated for \textit{overlapping pairs} of MLWFs, the number of Poisson equations that need to be solved is substantially decreased from ${\cal 
O}(N_b^2)$ to ${\cal O}(N_b)$. In addition,
$\overline{v}_{ij}(\rr)$ only needs to be solved on a subset of the
real-space grid (that is in general of fixed size) that encompasses
the overlap between a given pair of MLWFs. This further reduces the
overall computational effort required to evaluate exact-exchange
related quantities and results in a linear-scaling (${\cal O}(N_b)$)
algorithm. As such, this framework for performing exact-exchange
calculations is most efficient for non-metallic systems
(\ie, systems with a finite band gap) in which the occupied
KS orbitals can be efficiently localized. 

The MLWF representation not only yields the exact-exchange energy
$E_{\rm xx}$,
\begin{equation}
  E_{\rm xx} = -e^2 \sum_{ij} \int \dr \, \overline{\rho}_{ij}(\rr)
  \overline{v}_{ij}(\rr),
\label{eq:exx}
\end{equation}
at a significantly reduced computational
cost, but it also provides an amenable way of computing the
exact-exchange contributions to the (MLWF) wavefunction forces,
$\overline{D}_{xx}^{i}(\rr) = e^2 \sum_{j} \overline{v}_{ij}(\rr)
\overline{\varphi}_{j}(\rr)$, which serve as the central quantities in
Car-Parrinello MD simulations~\cite{Car:1985}. Moreover, the
exact-exchange contributions to the stress tensor are readily
available, thereby providing a general code base which enables hybrid
DFT based simulations in the NVE, NVT, and NPT ensembles for
simulation cells of any shape~\cite{Ko_2017}. We note in passing that
applications of the current implementation of this MLWF-based
exact-exchange algorithm are limited to $\Gamma$--point calculations
employing norm-conserving pseudo-potentials.  

\begin{figure}[ht]
\hbox to\hsize{\hfill
\includegraphics[width=0.45\textwidth]{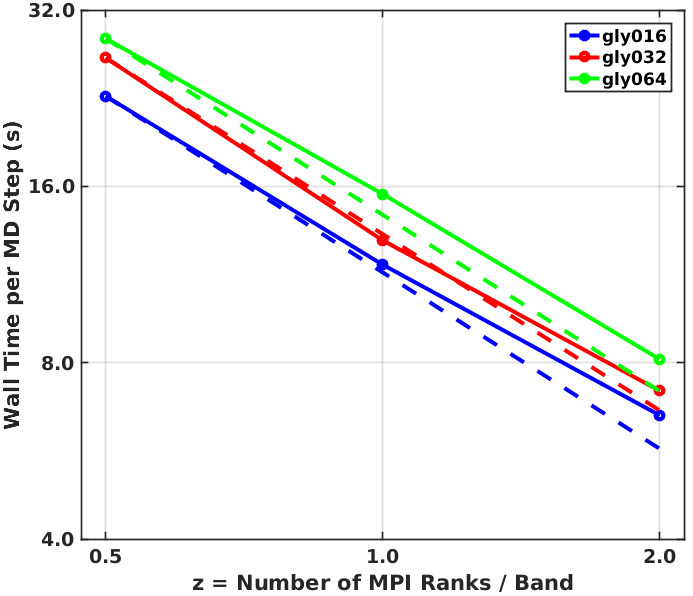} 
\hfill 
\includegraphics[width=0.45\textwidth]{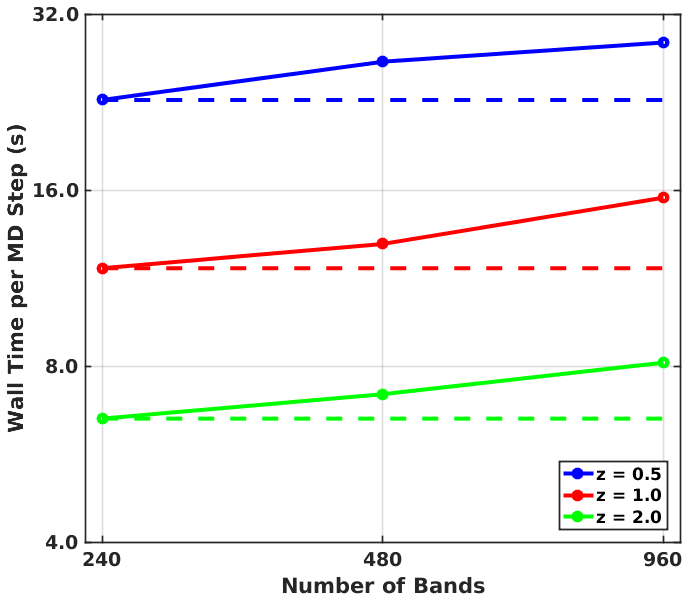}
\hfill}
\caption{\label{fig:scaling_mira} Strong (\textit{left}) and weak
  (\textit{right}) scaling plots on \textit{Mira} (BG/Q) for
  hybrid-DFT simulations of the $\alpha$-glycine molecular crystal
  polymorph using the linear-scaling exact-exchange algorithm in
  \CP. In these plots, unit cells containing 16-64 glycine molecules
  (160-640 atoms, 240-960 bands) were considered as a function of
  \var{z}, the number of MPI ranks per band (\var{z} = 0.5-2). On
  \textit{Mira}, 30,720 cores (1920 MPI ranks $\times$ 16 OpenMP
  threads/rank $\times$ 1 core/OpenMP thread) were utilized for the
  largest system (gly064, \var{z} = 2), retaining over 88\% (strong
  scaling) and 80\% (weak scaling) of the ideal efficiencies (dashed
  lines). Deviations from ideal scaling are primarily due to the FFT
  (which scales non-linearly) required to provide the MLWFs in real
  space.}  
\end{figure}

The MLWF-based exact-exchange algorithm in \CP\ employs a hybrid
MPI/OpenMP parallelization strategy that has been extensively
optimized for use on large-scale massively-parallel (super-) computer
architectures. The required set of Poisson equations---each one
treated as an independent task---are distributed across a large number
of MPI ranks/processes using a task distribution scheme designed to
minimize the communication and to balance computational
workload. Performance profiling demonstrates excellent scaling up to
30,720 cores (for the $\alpha$-glycine molecular crystal, see
Fig.~\ref{fig:scaling_mira}) and up to 65,536 cores (for
(H$_2$O)$_{256}$, see Ref. \onlinecite{DiStasio_2014}) on
\textit{Mira} (BG/Q) with extremely promising efficiency. In fact,
this algorithm has already been successfully applied to the study of
long-time MD simulations of large-scale condensed-phase systems such
as (H$_2$O)$_{128}$~\cite{DiStasio_2014,Santra_2015}. For more details
on the performance and implementation of this exact-exchange
algorithm, we refer the reader to Ref.~\onlinecite{Ko_2017}.  

\subsubsection{Dispersion interactions}

Dispersion, or van der Waals, interactions arise from dynamical
correlations  among charge fluctuations occurring in widely separated
regions of space. The resulting attraction is a non-local correlation
effect that cannot be reliably captured by any local (such as local
density approximation, LDA) or semi-local (generalized gradient
approximation, GGA) functional of the electron density
\cite{French_2010:long_range}. Such interactions can be either
accounted for by a truly non-local  exchange-correlation (XC)
functional, or modeled by effective interactions amongst atoms, whose
parameters are either computed from first  principles or estimated
semi-empirically. In \qe\ both approaches are implemented. Non-local
XC functionals are activated by selecting them in the \var{input\_dft}
variable, while explicit interactions are turned on with the
\var{vdw\_corr} option. From the latter group, DFT-D2
\cite{grimme2006}, Tkatchenko-Scheffler \cite{DFT-TS}, and
exchange-hole dipole moment models \cite{johnson2007,ncibook} are
currently implemented (DFT-D3 \cite{grimme2010} and the many-body
dispersion (MBD) \cite{MBD_2012,MBD_2014,MBD_2016} approaches are
already  available in a development version). 

\paragraph{Non-local van der Waals density functionals}
\label{sec:vdW-DF}
A fully non-local correlation functional able to account for van der
Waals interactions for general geometries was first developed in 2004
and named vdW-DF \cite{Dion_2004:van_waals}. Its development is firmly
rooted in many-body theory, where the so-called adiabatic connection
fluctuation-dissipation theorem (ACFD)
\cite{Langreth_1977:exchange-correlation_energy} provides a formally
exact expression for the XC energy through a coupling constant
integration over the response function---see Sec. \ref{sec:ACFD}. A
detailed review of the vdW-DF formalism is provided in
Ref.~\onlinecite{Berland_2015:van_waals}. The overall XC energy given
by the ACFD theorem---as a functional of the electron density $n$---is
then split in vdW-DF into a GGA-type XC part $E_{\rm xc}^0[n]$ and a
truly non-local correlation part $E_{\rm c}^{\rm nl}[n]$, \ie 
\begin{equation}
E_{\rm xc}[n] = E_{\rm xc}^0[n] + E_{\rm c}^{\rm nl}[n]\;,
\label{eq:splitup}
\end{equation}
where the non-local part is responsible for the van der Waals forces. Through a second-order expansion in the plasmon-response expression used to approximate the response function, the non-local part turns into a computationally tractable form involving a universal kernel $\Phi(\rr,\rr')$,
\begin{equation}
E_{\rm c}^{\rm nl}[n] = \frac{1}{2}\int \dr\;\dr'\;n(\rr)\;
\Phi(\rr,\rr')\;n(\rr')\;.
\label{eq:Ecnl}
\end{equation}
The kernel $\Phi(\rr,\rr')$ depends on $\rr$ and $\rr'$ only through $q_0(\rr)|\rr-\rr'|$ and $q_0(\rr')|\rr-\rr'|$, where $q_0(\rr)$ is a function of $n(\rr)$ and $\nabla n(\rr)$. As such, the kernel can be pre-calculated, tabulated, and stored in some external file. To make the scheme self-consistent, the XC potential $V_{\rm c}^{\rm nl}(\rr)=\delta E_{\rm c}^{\rm nl}[n]/\delta n(\rr)$ also needs to be computed \cite{Thonhauser_2007:van_waals}. The evaluation of $E_{\rm c}^{\rm nl}[n]$ in Eq.~(\ref{eq:Ecnl}) is computationally expensive. In addition, the evaluation of the corresponding potential  $V_{\rm c}^{\rm nl}(\rr)$ requires one spatial integral for each point $\rr$. A significant speedup can be achieved by writing the kernel in terms of splines \cite{Roman-Perez_2009:efficient_implementation}
\begin{eqnarray}
\Phi(\rr,\rr') &=& \Phi\big(q_0(\rr),q_0(\rr'),|\rr-\rr'|)\nonumber\\
&\approx& \sum_{\alpha\beta}\Phi(q_\alpha,q_\beta,|\rr-\rr'|)\;
p_\alpha\big(q_0(\rr)\big)\;p_\beta\big(q_0(\rr')\big)\;,
\end{eqnarray}
where $q_\alpha$ are fixed values and $p_\alpha$ are cubic splines. Equation~(\ref{eq:Ecnl}) then becomes a convolution that can be simplified to
\begin{eqnarray}
E_{\rm c}^{\rm nl}[n] &=& \frac{1}{2}\sum_{\alpha\beta}
\int \dr\;\dr'\;\theta_\alpha(\rr)\;
\Phi_{\alpha\beta}(|\rr-\rr'|)\;\theta_\beta(\rr')\nonumber\\
&=& \frac{1}{2}\sum_{\alpha\beta}\int \dk\;\theta_\alpha^*(\kk)\;
\Phi_{\alpha\beta}(k)\;\theta_\beta(\kk)\;.
\label{eq:Soler}
\end{eqnarray}
Here $\theta_\alpha(\rr)=n(\rr) p_\alpha\big(q_0(\rr)\big)$ and
$\theta_\alpha(\kk)$ is its Fourier transform. Accordingly,
$\Phi_{\alpha\beta}(k)$ is the Fourier transform of the original
kernel $\Phi_{\alpha\beta}(r)=\Phi(q_\alpha,q_\beta,|\rr-\rr'|)$.
Thus, two spatial integrals are replaced by one integral over Fourier
transformed quantities, resulting in a considerable speedup. This
approach also provides a convenient evaluation for $V_{\rm c}^{\rm
  nl}(\rr)$. 

The vdW-DF functional was implemented in \qe\ version 4.3, following
Eq.~(\ref{eq:Soler}). As a result, in large systems, compute times in
vdW-DF calculations are only insignificantly longer than for standard
GGA functionals.  The implementation uses a tabulation of the Fourier
transformed kernel $\Phi_{\alpha\beta}(k)$ from  Eq.~\eqref{eq:Soler}
that is computed by an auxiliary code,
\texttt{generate\_vdW\_kernel\_table.x}, and stored in the external
file \texttt{vdW\_kernel\_table}. The file then has to be placed
either in the directory where the calculation is run or in the
directory where the corresponding pseudopotentials reside. The
formalism for vdW-DF stress was derived and implemented in
Ref.~\onlinecite{Sabatini_2012:structural_evolution}.  The proper spin
extension of vdW-DF, termed svdW-DF
\cite{Thonhauser_2015:spin_signature}, was implemented in \qe\ version
5.2.1. 

Although the ACFD theorem provides guidelines for the total
XC functional in Eq.~(\ref{eq:splitup}), in
practice $E_{\rm xc}^0[n]$ is approximated by simple GGA-type
functional forms. This has been used to improve vdW-DF---and correct
the often too large binding separations found in its original
form---by optimizing the exchange contribution to $E_{\rm xc}^0[n]$.
The naming convention for the resulting variants is that the extension
should describe the exchange functional used. In this context, the
functionals vdW-DF-C09 \cite{Cooper_2010:van_waals}, vdW-DF-obk8
\cite{Klimes_2010:chemical_accuracy}, vdW-DF-ob86
\cite{Klimes_2011:van_waals}, and vdW-DF-cx
\cite{Berland_2014:exchange_functional} have been developed and
implemented in \qe. While all of these variants use the same kernel to
evaluate $E_{\rm c}^{\rm nl}[n]$, advances have also been made in
slightly adjusting the kernel form, which is referred to and
implemented as vdW-DF2 \cite{Lee_2010:higher-accuracy_van}. A
corresponding variant, \ie, vdW-DF2-b86r \cite{PhysRevB.82.153412},
is also implemented. Note that vdW-DF2 uses the same kernel file as
vdW-DF. 

The functional VV10 \cite{Vydrov_2010:nonlocal_van} is related to
vdW-DF, but adheres to fewer exact constraints and follows a very
different design philosophy. It is implemented in \qe\ in a form
called rVV10 \cite{Sabatini_2013:nonlocal_van} and uses a  different
kernel and kernel file that can be generated by running the auxiliary
code \exec{generate\_rVV10\_kernel\_table.x}. 

\paragraph{Interatomic pairwise dispersion corrections}

An alternative approach to accounting for dispersion forces is to add
to the XC energy $E^0_{\rm xc}$ a dispersion energy, $E_{\rm disp}$,
written as a damped asymptotic pairwise expression: 
\begin{equation}
  E_{\rm xc} = E^0_{\rm xc} + E_{\rm disp},\qquad
  E_{\rm disp} = - \frac{1}{2}\sum_{n=6,8,10}\sum_{I\neq J}
  \frac{C^{(n)}_{IJ}f_{n}(R_{IJ})}{R^{n}_{IJ}}
\label{eq:xdm}
\end{equation}
where $I$ and $J$ run over atoms, $R_{IJ}=|\RR_I-\RR_J|$ is the
interatomic distance between atoms $I$ and $J$, and $f_n(R)$ is a suitable
damping function. The interatomic dispersion coefficients
$C^{(n)}_{IJ}$ can be derived from fits, as in
DFT-D2~\cite{grimme2006}, or calculated non-empirically, as in the
Tkatchenko-Scheffler (TS-vdW)~\cite{DFT-TS} and exchange-hole dipole moment
(XDM) models~\cite{johnson2007,ncibook}. 

In XDM, the $C^{(n)}_{IJ}$ coefficients are calculated assuming that
dispersion interactions arise from the electrostatic attraction
between the electron-plus-exchange-hole distributions on different
atoms~\cite{johnson2007,ncibook}. In this way, XDM retains the
simplicity of a pairwise dispersion correction, like in DFT-D2, but
derives the $C^{(n)}_{IJ}$ coefficients from the electronic properties
of the system under study. The damping functions $f_n$ in Eq.~(\ref{eq:xdm}) suppress the
dispersion interaction at short distances, and serve the purpose of
making the link between the short-range correlation (provided by the
XC functional) and the long-range dispersion energy, as well as
mitigating erroneous behavior from the exchange functional in  the
representation of intermolecular repulsion~\cite{ncibook}. The damping
functions contain two adjustable parameters, available
online~\cite{schooner} for a number of popular density
functionals. Although any functional for which damping parameters are
available can be used, the functionals showing best performance when
combined with XDM appear to be B86bPBE~\cite{b86b,pbe} and
PW86PBE~\cite{pw86,pbe}, thanks to their accurate modeling of Pauli
repulsion~\cite{ncibook}. Both functionals have been implemented in
\qe\ since version 5.0. 

In the canonical XDM implementation, recently included in \qe\ and
described in detail elsewhere~\cite{xdmsolids}, the dispersion
coefficients are calculated from the electron density, its
derivatives, and the kinetic energy density, and assigned to the
different atoms in the system using a Hirshfeld atomic partition
scheme~\cite{hirshfeld}. This means that XDM is effectively a meta-GGA
functional of the dispersion energy whose evaluation cost is small
relative to the rest of the self-consistent calculation. Despite the
conceptual and computational simplicity of XDM, and because the
dispersion coefficients depend upon the atomic environment in a
physically meaningful way, the XDM dispersion correction offers good
performance in the calculation of diverse properties, such as lattice
energies, crystal geometries, and surface adsorption energies. XDM is
especially good for modeling organic crystals and organic/inorganic
interfaces. For a recent review, see Ref.~\citenum{ncibook}. 

The XDM dispersion calculation is turned on by specifying
\var{vdw\_corr='xdm'} and optionally selecting appropriate damping
function parameters (with the \var{xdm\_a1} and \var{xdm\_a2}
keywords). Because the reconstructed all-electron densities are
required during self-consistency, XDM can be used only in combination
with a PAW approach. The XDM contribution to forces and stress is not
entirely consistent with the energies because the current
implementation neglects the change in the dispersion
coefficients. Work is ongoing to remove this limitation, as well as to
make XDM available for Car-Parrinello MD, in future \qe\ releases. 

In the TS-vdW approach (\var{vdw\_corr='ts-vdw'}), all vdW parameters (which include the atomic dipole polarizabilities, $\alpha_I$, vdW radii, $R^0_I$, and interatomic $C^{(6)}_{IJ}$ dispersion coefficients) are functionals of the electron density and computed using the Hirshfeld partitioning scheme~\cite{hirshfeld} to account for the unique chemical environment surrounding each atom. This approach is firmly based on a fluctuating quantum harmonic oscillator (QHO) model and results in highly accurate $C^{(6)}_{IJ}$ coefficients with an 
associated error of approximately 5.5\% \cite{DFT-TS}. The TS-vdW approach requires a single empirical range-separation parameter based on the underlying XC functional and is recommended in conjunction with non-empirical DFT functionals such as PBE and PBE0. For a recent review of the TS-vdW approach and several other vdW/dispersion corrections, please see Ref.~\citenum{vdw-chem-rev_2017}.

The implementation of the density-dependent TS-vdW correction in \qe\ is fully self-consistent \cite{SCTSvdW_2015} and currently available for use with norm-conserving pseudo-potentials. An efficient linear-scaling implementation of the TS-vdW contribution to the ionic forces and stress tensor allows for Born-Oppenheimer and Car-Parrinello MD simulations at the DFT+TS-vdW level of theory; this approach has already been successfully employed in long-time MD simulations of large-scale condensed-phase systems such as (H$_2$O)$_{128}$ \cite{DiStasio_2014,Santra_2015}. We note in passing that the \qe\ implementation of the TS-vdW correction also includes analytical derivatives of the Hirshfeld weights, thereby completely reflecting the change in all TS-vdW parameters during geometry/cell optimizations and MD simulations.

\subsubsection{Hubbard-corrected functionals: DFT+U}
\label{sec:dft+U}

Most approximate XC functionals used in modern DFT codes fail quite
spectacularly on systems with atoms whose ground-state electronic 
structure features partially occupied, strongly localized orbitals 
(typically of the $d$ or $f$ kind), that suffer from strong 
self-interaction effects and a poor description of electronic correlations.
In these circumstances, DFT+U is often, although not always, an
efficient remedy. This method is based on the addition to the DFT
energy functional $E_\text{DFT}$ of a correction $E_U$, shaped on a
Hubbard model Hamiltonian: $E_{\text{DFT}+U} = E_{\text{DFT}} +
E_{U}$.  The original implementation in \qe, extensively described in
Refs. \onlinecite{cococcioni05, himmetoglu14}, is based on the
simplest rotationally invariant formulation of $E_U$, due to Dudarev
and  coworkers \cite{dudarev98}:
\begin{equation}
E_U = \frac{1}{2} \sum_IU^I\sum_{m,\sigma}\left\{ n^{I\sigma}_{mm} -
                          \sum_{m'} n^{I\sigma}_{mm'} n^{I\sigma}_{m'm}\right\},
\label{eq:Uener}
\end{equation}
where 
\begin{equation}
n^{I\sigma}_{mm'}= \sum_{\kk,\nu} f^\sigma_{\kk\nu} \langle\psi^\sigma_{\kk\nu}
 |\phi^I_m\rangle\langle \phi^I_{m'}| \psi^\sigma_{\kk\nu}\rangle,
\label{eq:Un}
\end{equation}
$|\psi^\sigma_{\kk\nu}\rangle$ is the valence electronic wave function for state
 $\kk\nu$ of spin $\sigma$, $f^\sigma_{\kk\nu}$ the corresponding occupation
 number, $|\phi^I_m\rangle$ is the chosen Hubbard manifold of atomic
 orbitals, centered on atomic site $I$, that may be orthogonalized or not.
 The presence of the Hubbard functional results in extra terms in
 energy derivatives such as forces, stresses,  elastic constants,
 or force-constant (dynamical) matrices. For instance, the additional term in
 forces is
 \begin{equation}
 {\bf F}^U_{I\alpha} = - \frac{1}{2} \sum_{I,m,m',\sigma}U^I\left(\delta_{mm'}
 -2 n^{I\sigma}_{mm'} \right)  \frac{\partial n^{I\sigma}_{mm'}}{\partial R_{I\alpha}}
\label{eq:Uforce}
\end{equation}
 where $R_{I\alpha}$ is the $\alpha$ component of position for atom $I$
 in the unit cell,
\begin{equation}
\frac{\partial n^{I\sigma}_{mm'}}{\partial R_{I\alpha}} =
\sum_{\kk,\nu} f^\sigma_{\kk\nu} \left( \left \langle\psi^\sigma_{\kk\nu}
\left |\frac{\partial \phi_{Im}}{\partial R_{I\alpha}}\right .\right  \rangle \langle \phi_{Im'}| \psi^\sigma_{\kk\nu}\rangle +
\langle \psi^\sigma_{\kk\nu} | \phi_{Im} \rangle
\left\langle \left . \frac{\partial \phi_{Im'}}{\partial R_{I\alpha}} \right | \psi^\sigma_{\kk\nu}\right \rangle\right).
\label{eq:Udn}
\end{equation}

\paragraph{Recent extensions of the formulation} 

As a correction to the total energy, the Hubbard functional naturally 
contributes an extra term to the total potential that enters 
the KS equations.
An alternative formulation \cite{sclauzero13}  of the DFT+U method,
recently introduced and implemented in \qe\
for transport calculations, eliminates the need of
extra terms in the potential by incorporating the Hubbard correction
directly into the (PAW) pseudopotentials through a renormalization of the
coefficients of their non-local terms.  

A simple extension to the Dudarev functional, DFT+U+J0,
was proposed in Ref. \citenum{himmetoglu11} and used to capture 
the insulating ground state of CuO. In CuO the localization of holes
on the $d$ states of Cu and the consequent on-set of a magnetic ground
state can only be stabilized against a competing tendency to hybridize
with oxygen $p$ states 
when on-site exchange interactions are precisely accounted for.
A simplified functional, depending upon the on-site (screened) Coulomb
interaction $U$ and the Hund's coupling $J$, can be obtained from the
full second-quantization
formulation of the electronic interaction potential
by keeping only on-site terms that describe the interaction between up to
two orbitals and by approximating on-site effective interactions with the 
(orbital-independent) atomic averages of Coulomb and exchange terms:
\begin{equation}
E_{U+J} = \sum_{I, \, \sigma}\, \frac{U^I - J^I}{2}\, 
{\rm Tr} \Bigl [ {\bf n}^{I\, \sigma}\, ( {\bf 1} - {\bf n}^{I\, \sigma} ) \Bigr ] + \sum_{I, \, \sigma}\, \frac{J^I}{2}\, {\rm Tr} \Bigl [ {\bf n}^{I\, \sigma}\, {\bf n}^{I\, -\sigma} \Bigr ]. 
\label{eq:dft_puj_1}
\end{equation}
The on-site exchange coupling $J^I$ not only reduces the effective Coulomb repulsion between like-spin electrons as in the simpler Dudarev functional (first term of the right-hand side), but also contributes a second term that acts as an extra penalty for the simultaneous presence of anti-aligned spins on the same atomic site and further stabilizes ferromagnetic ground states.

The fully rotationally invariant scheme of Liechtenstein {\it et al.} \cite{liechtenstein95}, generalized to non-collinear magnetism and two-component spinor wave-functions, is also implemented in the current version of \qe. The corrective energy term for each correlated atom can be quite generally written as: 
\begin{equation}
E_{U}^{full}=\frac{1}{2} \sum_{\alpha\beta\gamma\delta} U_{\alpha\beta\gamma\delta} \langle c_\alpha^\dagger c_\beta^\dagger c_\delta c_\gamma  \rangle_{\text{DFT}}=\frac{1}{2} \sum_{\alpha\beta\gamma\delta} \bigl ( U_{\alpha\beta\gamma\delta}-U_{\alpha\beta\delta\gamma} \bigr ) n_{\alpha\gamma}n_{\beta\delta}, 
\end{equation}
where the average is taken over the DFT Slater determinant, $U_{\alpha\beta\gamma\delta}$ are Coulomb integrals, and  some set of orthonormal spin-space atomic functions, $ \{\alpha\}$, is used to calculate the occupation matrix, $n_{\alpha\beta}$, via Eq.~(\ref{eq:Un}). These basis functions could be spinor wave functions of total angular momentum $j=l\pm 1/2$, originated from spherical harmonics of orbital momentum $l$, which is a natural choice in the presence of spin-orbit coupling. Another choice, adopted in our implementation, is to use the standard basis of separable atomic functions, $R_l(r)Y_{lm}(\theta,\phi)\chi(\sigma)$, where $\chi(\sigma)$ are spin up/down projectors and the radial function, $R_l(r)$, is an eigenfunction of the pseudo-atom. In the presence of spin-orbit coupling, the radial function is constructed by   averaging between the two radial functions $R_{l\pm1/2}$. These radial functions are read from the file containing the pseudopotential, in this case a fully-relativistic one. In this conventional basis, the corrective functional takes the form:
\begin{equation}
E_{U}^{full}=\frac{1}{2} \sum_{ijkl,\sigma\sigma'}
U_{ijkl}n_{ik}^{\sigma\sigma}n_{jl}^{\sigma'\sigma'}-
\frac{1}{2} \sum_{ijkl,\sigma\sigma'}
U_{ijlk}n_{ik}^{\sigma\sigma'}n_{jl}^{\sigma'\sigma},
\end{equation}
where $\{ijkl\}$ run over azimuthal quantum number $m$. 
The second term contains a spin-flip contribution if 
$\sigma'\ne \sigma$. For collinear magnetism, 
when $n_{ij}^{\sigma\sigma'}=\delta_{\sigma\sigma'}n_{ij}^{\sigma}$,
the present formulation reduces to the scheme
\cite{liechtenstein95} of Liechtenstein {\it et al.}
All Coulomb integrals, $U_{ijkl}$, can be parameterized by few input parameters such as $U$ ($s$-shell);
$U$ and $J$ ($p$-shell);
$U,J$ and $B$ ($d$-shell); 
$U, J, E_2$, and $E_3$ ($f$-shell), and so on.  
We note that if all parameters but $U$ are set to zero,
the Dudarev functional is recovered.

\paragraph{Calculation of Hubbard parameters} 

The Hubbard corrective functional $E_U$ depends linearly upon the 
effective on-site interactions, $U^I$. 
Therefore, using a proper value for these interaction parameters is crucial to 
obtain quantitatively
reliable results from DFT+U calculations. The \qe\ implementation of
DFT+U has also been the basis to develop 
a method for the calculation of $U$ \cite{cococcioni05}, based on
linear-response theory.
This method is completely ab initio 
and provides values of the effective interactions that are consistent with the system and with the ground state that the Hubbard functional aims at correcting.
A comparative analysis of this method with other approaches proposed in the 
literature to compute the Hubbard interactions has been initiated in
Ref.~\citenum{himmetoglu14}
and will be further refined in forthcoming publications
by the same authors.

Within linear-response theory, the Hubbard interactions are the elements of an effective
interaction matrix, computed as the difference between bare and screened inverse 
susceptibilities \cite{cococcioni05}: 
\begin{equation}
\label{eq:Ucalc}
U^I = \left(\chi_0^{-1} - \chi^{-1}\right)_{II}.
\end{equation}
In Eq.~(\ref{eq:Ucalc}) the susceptibilities $\chi$ and $\chi_0$ measure the
response of atomic occupations to shifts in the potential acting on the states 
of single atoms in the system. In particular, $\chi$ is defined as
$\chi_{IJ} = \sum_{m\sigma} \left(dn^{I\sigma}_{mm} / d\alpha^J\right) $
and is evaluated at self consistency, while  
$\chi_0$ has a similar definition but is computed before the self-consistent re-adjustment of the Hartree and XC potentials.
In the current implementation these susceptibilities are computed from a series of self-consistent
total energy calculations (varying the strength $\alpha$ of the perturbing potential over a range of values)
performed on supercells of sufficient size for the perturbations to be isolated from their
periodic replicas. While easy to implement, this approach
is quite cumbersome to use, requiring multiple calculations,
expensive convergence tests of the resulting parameters and complex post-processing tools.

These difficulties can be overcome by using 
density-functional perturbation theory (DFpT) 
to automatize the calculation of the Hubbard parameters.
The basic idea is to recast the entries of the susceptibility matrices
into sums over the BZ:
\begin{equation}
\label{eq:dnq}
\frac{dn^{I\sigma}_{mm'}}{d\alpha^J} = \frac{1}{N_\qq}\sum_\qq^{N_\qq} e^{i\qq\cdot(\RR_l - \RR_{l'})}\Delta_\qq^{s'} n^{s \,\sigma}_{mm'} \,,
\end{equation}
where $I\equiv(l,s)$ and $J\equiv(l',s')$, $l$ and $l'$ label
unit cells, $s$ and $s'$ label atoms in the unit cell,
$\RR_l$ and $\RR_{l'}$ are Bravais lattice vectors,
and $\Delta_\qq^{s'} n^{s \,\sigma}_{mm'}$
represent the (lattice-periodic) response of atomic occupations
to monochromatic perturbations constructed by modulating the shift
to the potential of all the periodic replica of a given atom by a
wave-vector $\qq$. This quantity is evaluated within
DFpT (see Sec. \ref{sec:lrexc}), using linear-response routines contained
in \texttt{LR\_Modules} (see Sec. \ref{sec:codereorg}).
This approach eliminates the need for supercell calculations in
periodic systems (along with the cubic scaling of their computational
cost) and automatizes complex post-processing operations needed to
extract $U$ from the output of calculations. The use of DFpT also offers
the perspective to directly evaluate inverse susceptibilities, thus
avoiding the matrix inversions of Eq.~(\ref{eq:Ucalc}), and to calculate
the Hubbard parameters for closed-shell systems, a notorious problem
for schemes based on perturbations to the potential. Full details
about this implementation will be provided in a forthcoming
publication \cite{Timrov:2017b} and the corresponding code
will be made available in one of the next \qe\ releases. 

\subsubsection{Adiabatic-connection fluctuation-dissipation theory}
\label{sec:ACFD}

In the quest for better approximations for the unknown
XC energy functional in KS-DFT, the  
approach based on the adiabatic connection fluctuation-dissipation
(ACFD) theorem~\cite{Langreth_1977:exchange-correlation_energy} has
received considerable interest in recent years. 
This is largely due to some attractive features: (i) a formally exact expression 
for the XC energy in term of density linear response functions can be 
derived providing a promising way for a systematic improvement of the XC 
functional; (ii) the method treats the exchange energy exactly, thus canceling 
out the spurious self-interaction error present in the Hartree energy; (iii) the 
correlation energy is fully non local and automatically includes long-range van 
der Waals interactions (see Sec. \ref{sec:vdW-DF}).

Within the ACFD framework a formally exact expression for the XC 
energy $E_{\rm xc}$ of an electronic system can be derived:
\begin{equation}
 E_{\rm xc} = -\frac{\hbar}{2 \pi} \int_{0}^{1}d\lambda{\int {\dr \dr'\frac{e^2}{|\rr-\rr'|}} \times \left [ \int_{0}^{\infty} 
{\chi_{\lambda}(\rr,\rr', iu) du + \delta(\rr-\rr') n(\rr)} \right ] }, 
\label{eq.Exc_acfdt}
\end{equation}
where $\hbar=h/2\pi$ and $h$ is the Planck constant, 
$\chi_{\lambda}(\rr,\rr';iu)$ is the density response function 
at imaginary frequency $iu$ of a system whose electrons interact via a
scaled Coulomb interaction, \ie, $\lambda e^2/|\rr-\rr'|$, and are
subject to a local potential such that the electronic density $n(\rr)$
is independent of $\lambda$, and is thus equal to the ground-state
density of the fully interacting system $(\lambda=1)$. The XC energy,  
Eq.~(\ref{eq.Exc_acfdt}), can be further separated into a KS exact-exchange 
energy $E_{\rm xx}$, Eq.~(\ref{eq:exx}), and a correlation energy $E_{\rm c}$.
The former 
is routinely evaluated as in any hybrid functional calculation
(see Sec.~\ref{sec:exx}).
Using a matrix notation, the latter can be expressed in a compact 
form in terms of the Coulomb interaction, $v_c = e^2/|\rr-\rr'|$, 
and of the density response functions:
\begin{equation}
 E_{\rm c} = -\frac{\hbar}{2 \pi} \int_{0}^{1} d\lambda \int_{0}^{\infty} du\; \rm{tr} \bigl [ v_c [\chi_{\lambda}(iu) 
-\chi_{0}(iu)] \bigr ] . 
\label{eq:ec_acfdt}
\end{equation}
For $\lambda > 0$, $\chi_{\lambda}$ can be related to the noninteracting 
density response function $\chi_{0}$ via a Dyson equation obtained from 
TDDFT:
\begin{equation}
 \chi_{\lambda}(iu) = \chi_0(iu) + \chi_0(iu) \bigl [\lambda v_c + f_{\rm xc}^{\lambda} (iu) \bigr ] \chi_{\lambda}(iu).
\end{equation}
The exact expression of the XC kernel $f_{\rm xc}$ is unknown, and in 
practical applications one needs to approximate it. 
In the \texttt{ACFDT} package, the random phase approximation (RPA), 
obtained by setting $f_{\rm xc}^{\lambda} = 0$, and the RPA plus 
exact-exchange kernel (RPAx), obtained by setting $f_{\rm xc}^{\lambda}= 
\lambda f_{\rm x}$, are implemented. 
The evaluation of the RPA and RPAx correlation energies is based on an 
eigenvalue decomposition of the non-interacting response functions and of 
its first-order correction in the limit of vanishing electron-electron 
interaction~\cite{PhysRevB.78.113303,Nguyen-HV:2009,Colonna:2014}. Since
only a small number 
of these eigenvalues are relevant for the calculation of the correlation 
energy, an efficient iterative scheme can be 
used to compute the low-lying modes of the RPA/RPAx density 
response functions. 

The basic operation required for the eigenvalue decomposition is a 
number of loosely coupled DFpT calculations for different imaginary 
frequencies and trial potentials. Although the global scaling of the
iterative approach is the same as for implementations
based on the evaluation of the full response matrices ($N^4$), 
the number of operation involved is 100 to 1000 times
smaller~\cite{Nguyen-HV:2009}, thus largely reducing the 
global scaling pre-factor.
Moreover, the calculation can be parallelized very
efficiently by distributing different trial potentials on different 
processors or groups of processors.

In addition, the local EXX and RPA-correlation potentials can be computed
through an optimized effective potential (OEP) scheme fully compatible with
the eigenvalue decomposition strategy adopted for the evaluation of the 
EXX/RPA energy. Iterating the energy and the OEP calculations 
and using an effective mixing scheme to update the KS potential, a 
self-consistent minimization of the EXX/RPA functional can be
achieved~\cite{Nguyen-NL:2014}.

\subsection{Linear response and excited states without virtual orbitals}
\label{sec:lrexc}

One of the key features of modern DFT implementations is that they do
not require the calculation of virtual (unoccupied) orbitals. This
idea, first pioneered by Car and Parrinello in their landmark 1985
paper \cite{Car:1985} and later adopted by many groups world-wide,
found its way in the computation of excited-state properties with the
advent of density-functional perturbation theory (DFpT)
\cite{Baroni:1987,Giannozzi:1991,Gonze:1995,Baroni:2001}. DFpT is
designed to deal with static perturbations and its use is therefore
restricted to those excitations that can be described in the
Born-Oppenheimer approximation, such as lattice vibrations. The main
idea underlying DFpT is to represent the linear response of KS
orbitals to an external perturbation as generic orbitals satisfying an
orthogonality constraint with respect to the occupied-state manifold
and a self-consistent Sternheimer equation
\cite{Sternheimer:1954,Mahan:1980}, rather than as linear combinations
of virtual orbitals (which would require the computation of all, or a
large number, of them).

Substantial progress has been made over the past decade, allowing one to extend DFpT to the dynamical regime, and thus simulate sizable portions of the optical and loss spectra of complex molecular and extended systems, without making any explicit reference to their virtual states. Although the Sternheimer approach can be easily extended to time-dependent perturbations \cite{Schwartz:1959,Zernik:1964,Baroni:1985}, its use is hampered in practice by the fact that a different Sternheimer equation has to be solved for each different value of the frequency of the perturbation. When the perturbation acting on the system vanishes, the frequency-dependent Sternheimer equation becomes a non-Hermitian eigenvalue equation, whose eigenvalues are the excitation energies of the system. In the TDDFT community, this equation is known as the Casida equation \cite{Casida:1996,Jamorski:1996}, which is the immediate translation to the DFT parlance of the time-dependent Hartree-Fock equation \cite{McLachlan1964}. This approach to excited states is optimal in those cases where one is interested in a few excitations only, but can hardly be extended to continuous spectra, such as those arising in extended systems or above the ionization threshold of even finite ones. In those cases where extended portions of a continuous spectrum is required, a new method has been developed, based on the Lanczos (bi-) orthogonalization algorithm, and dubbed the Liouville-Lanczos approach to time-dependent density-functional perturbation theory (TDDFpT). This method allows one to reuse intermediate products of an iterative process, essentially identical to that used for static perturbations, to build dynamical response functions from which spectral properties can be computed for a whole wide spectral range at once \cite{Walker:2006,Rocca:2008}.  A similar approach to linear optical spectroscopy was proposed later, based on the multi-shift conjugate gradient algorithm \cite{Hubener2014}, instead of Lanczos.   This powerful idea has been generalized to the solution of the Bethe-Salpeter equation, which is formally very similar to the eigenvalue equations arising in TDDFpT \cite{Rocca:2010,Rocca:2012,Marsili:2017}, and to the computation of the polarization propagator and self-energy operator appearing in the $GW$ equations \cite{Umari:2009,Umari:2010,Govoni:2015}. It is presently exploited in several components of the \qe\ distribution, as well as in other advanced implementations of many-body perturbation theory \cite{Govoni:2015}.  

\subsubsection{Static perturbations and vibrational spectroscopy}

The computation of vibrational properties in extended systems is one of the traditional fields of application of DFpT, as thoroughly described, \eg, in Ref.~\onlinecite{Baroni:2001}. The latest releases of \qe\ feature the linear-response implementation of several new functionals in the van der Waals and DFT+U families. Explicit expressions of the XC kernel, implementation details, and a thorough benchmark are reported in Ref.~\onlinecite{PhysRevB.93.235120} for the first case. 
As for the latter, DFpT+U has been implemented for both the Dudarev
\cite{dudarev98} and DFT+U+J0 functionals \cite{himmetoglu11},
allowing one to account for electronic localization effects acting
{\it selectively} on specific phonon modes at  arbitrary wave-vectors,
thus substantially improving the description of the
vibrational properties of strongly correlated systems with respect to
``standard'' LDA/GGA functionals. The current implementation allows
for both norm-conserving and ultrasoft pseudopotentials, insulators
and metals alike, also including the spin-polarized case. The
implementation of DFpT+U requires two main additional ingredients with
respect to standard DFpT \cite{floris11}.  
First, the dynamical matrix contains an additional term coming from the second derivative of the Hubbard term $E_U$ with respect to the atomic  positions (denoted $\lambda$ or $\mu$), namely:
\begin{equation}
\Delta^{\mu}(\partial^{\lambda} E_U) = \sum_{I\sigma m m'} U^I
\left[\frac{\delta_{mm'}}{2} - n^{I\sigma}_{mm'}\right] 
\Delta^{\mu}\left(\partial^{\lambda} n^{I\sigma}_{mm'}\right) 
-\sum_{I\sigma m m'} U^I \Delta^{\mu}n^{I\sigma}_{mm'}\partial^{\lambda}n^{I\sigma}_{mm'},
\label{eq:d2e}
\end{equation}
\normalsize
where the notations are the same as in Eq.~(\ref{eq:Uener}). The symbols $\partial$ and  $\Delta$ indicate, respectively, a {\it bare} derivative (leaving the KS wavefunctions unperturbed) and a {\it total} derivative (including also linear-response contributions). Second, in order to obtain a consistent electronic density response to the atomic displacements from the DFT+U  ground state, the  perturbed KS potential $\Delta V_{SCF}$ in the Sternheimer equation is augmented with the Hubbard perturbed potential $\Delta^{\lambda} V_U$: 
\begin{multline}
\qquad \Delta^{\lambda} V_U =
\sum_{I\sigma mm'} U^{I}
\left[\frac{\delta_{mm'}}{2}-n^{I\sigma}_{mm'}\right]  
\times \left[ |\Delta^{\lambda} \phi^I_{m'}\rangle \langle\phi^I_m| + |\phi^I_{m'}\rangle \langle \Delta^{\lambda}\phi^I_m| \right] \\ 
- \sum_{I\sigma mm'}U^{I} \Delta^{\lambda} n^{I\sigma}_{mm'}|\phi^I_{m'}\rangle
\langle\phi^I_m|,  \qquad
\label{eq:dvu}
\end{multline}
where the notations are the same as in Eq.~(\ref{eq:Un}). The unperturbed Hamiltonian in the Sternheimer equation  is the DFT+U Hamiltonian (including the Hubbard potential $V_U$). More implementation details will be given in a forthcoming publication~\cite{floris17}. 

Applications of DFpT+U include the calculation of the vibrational spectra of transition-metal monoxides like MnO and NiO \cite{floris11}, investigations of properties of materials of geophysical interest like goethite \cite{blanchard14, blanchard14bis}, iron-bearing \cite{shukla15,shukla16} and aluminum-bearing bridgmanite \cite{shukla16bis}. These results feature a significantly better agreement with experiment of the predictions of various lattice-dynamical properties, including the LO-TO and magnetically-induced TO splittings, as compared with standard LDA/GGA calculations.

\subsubsection{Dynamic perturbations: optical, electron energy loss, and magnetic spectroscopies} \label{sec:TDDFPT}

Electronic excitations can be described in terms of the dynamical charge- and spin-density susceptibilities, which are accessible to TDDFT \cite{Runge:1984,Marques:2012}.   In the linear regime the TDDFT equations can be solved using first-order perturbation theory. The time Fourier transform of the charge-density response, $\tilde{n}'(\rr,\omega)$, is determined by the projection over the unoccupied-state manifold of the Fourier transforms of the first-order corrections to the one-electron orbitals, $\tilde{\psi}'_{\kk\nu}(\rr,\omega)$, \cite{Walker:2006, Rocca:2008, Malcoglu:2011, Ge:2014, Baroni:2012}. For each band index $(\kk\nu)$, two response orbitals $x_{\kk\nu}$ and  $y_{\kk\nu}$ can be defined as  
\begin{eqnarray}
x_{\kk\nu}(\rr) &=& \frac1{2} \hat{Q} \left(\tilde{\psi}'_{\kk\nu}(\rr,\omega) + 
\tilde{\psi}'^*_{-\kk\nu}(\rr,-\omega)\right) \label{eq:change_variables_batch_1} \\
y_{\kk\nu}(\rr) &=& \frac1{2} \hat{Q} \left(\tilde{\psi}_{\kk\nu}(\rr,\omega) - 
\tilde{\psi}'^*_{-\kk\nu}(\rr,-\omega)\right),
\label{eq:change_variables_batch_2}
\end{eqnarray}
where $\hat{Q}$ is the projector on the unoccupied-state manifold. The
response orbitals $x_{\kk\nu}$ and $y_{\kk\nu}$ can be collected in
so-called {\em batches} $X=\{x_{\kk\nu}\}$ and $Y=\{y_{\kk\nu}\}$,
which uniquely determine the response density matrix. In a similar
way, the perturbing potential $\hat{V}'$ can be represented by the
batch $Z=\{z_{\kk\nu}\}=\{\hat{Q} \hat V' \psi_{\kk\nu}\}$. Using
these definitions, the linear-response equations of TDDFpT take the
simple form:  
\begin{equation}
\left( \hbar\omega - \hat{\mathcal L}\right) \cdot \left(
\begin{array}{c} X\\Y \end{array} \right) =
\left( \begin{array}{c} 0\\ Z \end{array}
\right), \qquad
\hat{\mathcal L} =
\left(
\begin{array}{cc}
0 & \hat{D}\\
\hat{D}+\hat{K} & 0
\end{array} \right),
\label{eq:Liouville}
\end{equation}
where the super-operators $\hat{D}$ and $\hat{K}$, which enter the definition of the {\em Liouvillian} super-operator, $\hat{\mathcal L}$, are defined in terms of the unperturbed Hamiltonian and of the perturbed Hartree-plus-XC potential \cite{Walker:2006, Rocca:2008, Malcoglu:2011, Ge:2014, Baroni:2012}. This implies that a Liouvillian build costs roughly twice as much as a single iteration in time-independent DFpT. It is important to note that $\hat{D}$ and $\hat{K}$, and therefore $\hat{\mathcal{L}}$, do not depend on the frequency $\omega$. For this reason, when in Eq.~(\ref{eq:Liouville}) the vector on the right-hand side, $(0,Z)^\top$, is set to zero, a linear eigenvalue equation is obtained (Casida's equation). 

The quantum Liouville equation~\eqref{eq:Liouville} can be seen as the
equation for the response density matrix operator
$\hat{\rho}'(\omega)$, namely $(\hbar\omega - \hat{\mathcal L}) \cdot
\hat{\rho}'(\omega) = [\hat{V}', \hat{\rho}^\circ]$, where
$[\cdot,\cdot]$ is the commutator and $\hat{\rho}^\circ$ is the
ground-state density matrix operator. With this at hand, we can
define a generalized susceptibility $\chi_{AV}(\omega)$, which
characterizes the response of an arbitrary one-electron Hermitian
operator $\hat{A}$ to the external perturbation $\hat{V}'$ as  
\begin{equation}
\chi_{AV}(\omega) = \mathrm{Tr}\left[ \hat{A} \hat{\rho}'(\omega) \right]
= \left\langle \hat{A} \, \left| \, (\hbar \omega - \hat{\mathcal L})^{-1} \right. \cdot [\hat{V}', \hat{\rho}^\circ] \right\rangle , \label{eq:chi-AV}
\end{equation}
where $\langle\cdot|\cdot\rangle$ denotes a scalar product in operator space. For instance, when both $\hat{A}$ and $\hat{V}'$ are one of the three Cartesian components of the dipole (position) operator, Eq.~\eqref{eq:chi-AV} gives the dipole polarizability of the system, describing optical absorption spectroscopy \cite{Walker:2006,Rocca:2008}; setting $\hat{A}$ and $\hat{V}'$ to one of the space Fourier components of the electron charge-density operator would correspond to the simulation of electron energy loss or inelastic X-ray scattering spectroscopies, giving access to plasmon and exciton excitations in extended systems \cite{Timrov:2013,Timrov:2015}; two different Cartesian components of the Fourier transform of the spin polarization would give access to spectroscopies probing magnetic excitations (\eg\ inelastic neutron scattering or spin-polarized electron energy loss) \cite{Gorni:2017}, and so on. When dealing with macroscopic electric fields, the dipole operator in periodic boundary conditions is treated using the standard DFpT prescription, as explained in Refs.~\cite{Baroni:1986,Tobik:2004}. 

The \qe \ distribution contains several codes to solve the Casida's equation or to directly compute generalized susceptibilities according to Eq.~\eqref{eq:chi-AV} and by solving Eq.~(\ref{eq:Liouville}) using different approaches for different pairs of $\hat{A}$/$\hat{V}'$, corresponding to different spectroscopies.
In particular, Eq.~\eqref{eq:Liouville} can be solved iteratively using the Lanczos recursion algorithm, which allows one to avoid computationally expensive inversion of the Liouvillian.
The basic principle of how matrix elements of the resolvent of an
operator can be calculated using a Lanczos recursion chain has been
worked out by Haydock, Heine, and Kelly~\cite{Haydock1972,Haydock1975}
for the case of Hermitian operators and diagonal matrix elements. The
quantity of interst can be written as
\begin{equation}
g_v(\omega) = \left\langle v \left| (\hbar\omega - \hat{L})^{-1}
\right. v \right\rangle \,.
\label{resolvent}
\end{equation}
A chain of vectors is defined by
\begin{eqnarray}
|q_0\rangle &=& 0 \label{rec1}\\
|q_1\rangle &=& |v\rangle \label{rec2}\\
\alpha_n &=& \langle q_n | \hat{L} q_n \rangle \label{rec3}\\
\beta_{n+1} \, |q_{n+1}\rangle &=& (\hat{L} - \alpha_n) \,
|q_n\rangle - \beta_{n} \, |q_{n-1}\rangle \,,
\label{rec4}
\end{eqnarray}
where $\beta_{n+1}$ is given by the condition $\langle q_{n+1}|q_{n+1}
\rangle = 1$. The vectors $|q_n\rangle$ created by this recursive
chain are orthonormal. Furthermore, the operator $\hat{L}$, written in the
basis of these vectors, is tridiagonal. If one limits the chain to the
$M$ first vectors $|q_0\rangle, |q_1\rangle,\cdots,|q_M\rangle$, then
the resulting representation of $\hat{L}$ is a $M\times M$ square matrix
$T_M$ which reads
\begin{equation}
T_M = \left(
\begin{array}{ccccc}
\alpha_1 & \beta_2 & 0 & \cdots & 0\\
\beta_2 & \alpha_2 & \beta_3 & \ddots & \vdots \\
0 & \beta_3 & \ddots & \ddots & 0\\
\vdots & \ddots & \ddots & \alpha_{M-1} & \beta_{M} \\
0 & \cdots & 0 & \beta_{M} & \alpha_M
\end{array}\right).
\label{tridiagonal}
\end{equation}
Using such a truncated representation of $\hat{L}$, the resolvent in
Eq.~(\ref{resolvent}) can be approximated as
\begin{equation}
g_v(\omega) \approx \left\langle v \left|\left( \hbar\omega - T_M \right)^{-1} \right. v \right\rangle \,.
\end{equation}
Thanks to the tridiagonal form of $T_M$, the approximate resolvent
can finally be written as a continued fraction
\begin{equation}
g_v(\omega) \approx \frac{1}{\hbar\omega - \alpha_1 +
  \frac{\displaystyle\beta_2^2}{\displaystyle \hbar\omega - \alpha_2 + ...}} \,.
\label{continuedfraction}
\end{equation}
Note that performing the recursion \eqref{rec1} -- \eqref{rec4} is 
the computational bottleneck of this algorithm, 
while evaluating the continued fraction in
Eq.~(\ref{continuedfraction}) is very fast. The recursion being
independent of the frequency $\omega$, a single
recursion chain yields information about any desired number of
frequencies, at negligible additional computational cost.
It is also important to note that at any stage of the recursion chain,
only three vectors need to be kept in memory, namely
$|q_{n-1}\rangle$, $|q_{n}\rangle$, and $|q_{n-1}\rangle$. This is a
considerable advantage with respect to the direct calculation of $N$
eigenvectors of $\hat{L}$ where all $N$ vectors need to be kept in memory in
order to enforce orthogonality.

The Liouvillian $\hat{\cal L}$ in Eq.~(\ref{eq:Liouville}) is not a
Hermitian operator. For this reason, the Lanczos algorithm presented
above cannot be directly applied to the calculation
of the generalized susceptibility~\eqref{eq:chi-AV}. There are two distinct
algorithms that can be applied in the non-Hermitian case. The
{\em non-Hermitian Lanczos biorthogonalization algorithm}~\cite{Rocca:2008,Malcoglu:2011} 
amounts to recursively applying the operator $\hat{\cal L}$ and it
Hermitian conjugate $\hat{\cal L}^{\dag}$ to {\em two} previous Lanczos
vectors $|v_n\rangle$ and $|w_n\rangle$. In this way, a pair of
bi-orthogonal basis sets is created. The operator $\hat{\cal L}$ can then
be represented in this basis as a tridiagonal matrix, similarly to
the Hermitian case, Eq.~(\ref{tridiagonal}).
The Liouvillian $\hat{\cal L}$ of TDDFT belongs to a special class of
non-Hermitian operators which are called
pseudo-Hermitian~\cite{Gruning:2011,Ge:2014}. For such operators, there
exists a second recursive algorithm to compute the resolvent --
{\em pseudo-Hermitian Lanczos algorithm} -- which is
numerically more stable and requires only half the numbers of
operations per Lanczos step~\cite{Gruning:2011,Ge:2014}.
Both algorithms have been implemented in \qe. Because of its speed and numerical stability, 
the use of the pseudo-Hermitian method is recommended.
 
This methodology has also been extended---presently only in the case of absorption spectroscopy---to employ hybrid functionals \cite{Rocca:2010,Rocca:2012,Ge:2014} (see Sec. \ref{sec:exx}). In this case the calculation requires the evaluation of the linear response of the non-local Fock potential, which is readily available from the response density matrix, represented by the \emph{batches} of response orbitals. The corresponding hybrid-functional Liouvillian features additional terms with respect to the definition in Eq.~\eqref{eq:Liouville}, but presents a similar structure and similar mathematical properties. Accordingly, semi-local and hybrid-functional TDDFpT employ the same numerical algorithms in practical calculations. 

\paragraph{Optical absorption spectroscopy} \label{sec:optical} The \exec{turbo\_lanczos.x} \cite{Malcoglu:2011, Ge:2014} and \exec{turbo\_davidson.x} \cite{Ge:2014} codes are designed to simulate the optical response of molecules and clusters. \exec{turbo\_lanczos.x} computes the dynamical dipole polarizability [see Eq.~\eqref{eq:chi-AV}] of finite systems over extended frequency ranges without ever computing any eigenpairs of the Casida equation. This goal is achieved by applying a recursive non-Hermitian or pseudo-Hermitian Lanczos algorithm. The two flavours of the Lanczos algorithm implemented in \exec{turbo\_lanczos.x} are particularly suited in those cases where one is interested in the spectrum over a wide frequency range comprising a large number of individual excitations. In \exec{turbo\_davidson.x} a Davidson-like algorithm \cite{Davidson:1975} is used to selectively compute a few eigenvalues and eigenvectors of ${\hat{\mathcal L}}$. This is useful when very few low-lying excited states are needed and/or when the excitation eigenvector is explicitly needed, \eg, to compute ionic forces on excited potential energy surfaces, a feature that will be implemented in one of the forthcoming releases. Both \exec{turbo\_lanczos.x} and \exec{turbo\_davidson.x} are interfaced with the \Environ\ module~\cite{Andreussi:2012}, to simulate the absorption spectra of complex molecules in solution using the self-consistent continuum solvation model \cite{Timrov:2015b} (see Sec.~\ref{sec:SCCS}). 

\paragraph{Electron energy loss spectroscopy}
\label{sec:eels}
The \exec{turbo\_eels.x} code \cite{Timrov:2015} computes the response of extended systems to an incoming beam of electrons or X rays, aimed at simulating electron energy loss (EEL) or inelastic X-ray scattering (IXS) spectroscopies, sensitive to collective charge-fluctuation excitations, such as plasmons. Similarly to the description of vibrational modes in a lattice by the \PHonon\ package, here the perturbation can be represented as a sum of monochromatic components corresponding to different momenta, $\qq$, and energy transferred from the incoming electrons to electrons of the sample. The quantum Liouville equation~\eqref{eq:Liouville} in the batch representation can be formulated for individual $\qq$ components of the perturbation, which can be solved independently \cite{Timrov:2013}. The recursive Lanczos algorithm is used to solve iteratively the quantum Liouville equation, much like in the case of absorption spectroscopy. The entire EEL/IXS spectrum is obtained in an arbitrarily wide energy range (up to the core-loss region) with only one Lanczos chain. Such a numerical algorithm allows one to compute directly the diagonal element of the charge-density susceptibility, see Eq.~\eqref{eq:chi-AV}, by avoiding computationally expensive matrix inversions and multiplications characteristic of standard methods based on the solution of the Dyson equation \cite{orr02}. The current version of \exec{turbo\_eels.x} allows to explicitly account for spin-orbit coupling effects \cite{Timrov:2017}.  

\paragraph{Magnetic spectroscopy}
\label{sec:magnons}
The response of the system to a magnetic perturbation is described by a spin-density susceptibility matrix, see Eq.~\eqref{eq:chi-AV}, labeled by the Cartesian components of the perturbing magnetic field and magnetization response, whose poles characterize spin-wave (magnon) and Stoner excitations. The methodology implemented in \exec{turbo\_eels.x} to deal with charge-density fluctuations has been generalized to spin-density fluctuations so as to deal with magnetic (spin-polarized neutron and electron) spectroscopies in extended systems. In the spin-polarized formulation of TDDFpT the time-dependent KS wave functions are two-component spinors $\{\psi^\sigma_{\kk\nu}(\rr,t)\}$ ($\sigma$ is the spin index), which satisfy a time-dependent Pauli-type KS equations and describe a time-dependent spin-charge-density, $n_{\sigma \sigma'}(\rr,t) = \sum_{\kk\nu} \psi^{\sigma *}_{\kk\nu}(\rr,t) \psi^{\sigma'}_{\kk\nu}(\rr,t)$. Instead of using the latter quantity it is convenient to change variables and use the charge density $n(\rr,t) = \sum_\sigma n_{\sigma \sigma}(\rr,t)$ and the spin density $\mathbf{m}(\rr,t) = \mu_B \sum_{\sigma \sigma'} \boldsymbol{\sigma}_{\sigma \sigma'} \, n_{\sigma' \sigma}(\rr,t)$, where $\mu_B$ is the Bohr magneton and $\boldsymbol{\sigma}$ is the vector of Pauli matrices. In the linear-response regime, the charge- and spin-density response $n'(\rr,t)$ and $\mathbf{m}'(\rr,t)$ are coupled via the scalar and magnetic XC response potentials $V'_{\rm xc}(\rr,t)$ and $\mathbf{B}'_{\rm xc}(\rr,t)$, which are treated on a par with the Hartree response potential $V'_\mathrm{H}(\rr,t)$, depending only on $n'(\rr,t)$, and which all enter the linear-response time-dependent Pauli-type KS equations. The lack of time-reversal symmetry in the ground state means that the TDDFpT equations have to be generalized to treat KS spinors at $\kk$ and $-\kk$ and various combinations with the $\qq$  vector. Moreover, this also implies that no rotation of batches is possible, as in Eqs.~\eqref{eq:change_variables_batch_1} and \eqref{eq:change_variables_batch_2}, and a generalization of the Lanczos algorithm to complex arithmetics is required. At variance with the cumbersome Dyson's equation approach, which requires the separate calculation and coupling of charge-charge, spin-spin, and charge-spin independent-electron polarizabilities, in our approach the coupling between spin and charge fluctuations is naturally accounted for via Lanczos chains for the spinor KS response orbitals. The current implementation supports general non-collinear spin-density distributions, which allows us to account for spin-orbit interaction and magnetic anisotropy. 
All the details about the present formalism will be given in a forthcoming publication \cite{Gorni:2017} and the corresponding code will be made available in one of the next \qe\ releases. 
\subsubsection{Many-body perturbation theory}

Many-body perturbation theory refers to a set of computational methods, based on quantum field theory, that are designed to calculate electronic 
and optical excitations beyond standard DFT~\cite{orr02}. 
The most popular among such methods are the $GW$ approximation and the 
Bethe-Salpeter equation (BSE) approach. The former is intended to calculate accurate quasiparticle
excitations, \eg, ionization energies and electron affinities in molecules,
band structures in solids, and accurate band gaps in semiconductor and insulators.
The latter is employed to study optical excitations by including electron-hole interactions.

In the $GW$ method the XC potential of DFT is corrected using a
many-body self-energy consisting of the product of the electron
Green's function  $G$ and the screened Coulomb interaction
$W$~\cite{hedin65, hl85}, which represents the lowest-order term in
the diagrammatic expansion of the exact electron self-energy. In the
vast majority of $GW$ implementations, the evaluation of $G$ and $W$
requires the calculation of both occupied and unoccupied KS
eigenstates. The convergence of the resulting self-energy correction
with respect to the number of unoccupied states is rather slow, and in
many cases it constitutes the main bottleneck in the
calculations. During the past decade there has been a growing interest
in alternative techniques which only require the calculation of
occupied electronic states, and several computational strategies have
been developed \cite{rog97, wlgg09, Umari:2010, gcl10}. The common
denominator to all these strategies is that they rely on
linear-response DFpT and the Sternheimer equation, as in the
\PHonon\ package. 

In \qe\ the $GW$ approximation is realized based
on a DFpT representation of response and self-energy operators, thus
avoiding any explicit reference to unoccupied states.
There are two different implementations: the
\GWL\ ($GW$-Lanczos) package \cite{Umari:2009,Umari:2010} and the
\SternheimerGW\ package \cite{sternheimergw}. The former focuses on
efficient $GW$ calculations for large systems (including  disordered
solids, liquids, and interfaces), and also supports the calculations
of optical spectra via the Bethe-Salpeter approach
\cite{Marsili:2017}. The latter focuses on high-accuracy  calculations
of band structures, frequency-dependent self-energies, and
quasi-particle spectral functions for crystalline solids. In addition
to these, the WEST code \cite{Govoni:2015}, not part of the
\qe\ distribution, relies on \qe\ as an external library to
perform similar tasks and to achieve similar goals. 

\paragraph{\GWL}
The  \GWL\  package consists of four different codes. The
\exec{pw4gww.x} code reads the KS wave-functions and charge density
previously calculated by \PWscf\  and prepares a set  of data which
are used by code \exec{gww.x} to perform the actual $GW$
calculation. While  \exec{pw4gww.x} uses the  plane-wave
representation of orbitals and charges  and the same \qe\  environment
as all other linear response codes, \exec{gww.x}  does not rely on any
specific representation of the orbitals. Its parallelization strategy
is based on the distribution of frequencies. Only a few basic
routines, such as the MPI drivers, are common with the rest of \qe. 

\GWL\ supports three different basis sets for representing polarisability operators: \emph{i)} plane wave-basis set, defined by an energy cutoff; \emph{ii)} the basis set formed by the most important eigenvectors (\ie, corresponding to the highest eigenvalues) of the actual irreducible polarisability operator at zero frequency calculated through linear response; \emph{iii)} the basis set formed by the  most important eigenvectors of an approximated  polarisability operator. The last  choice permits the control of the balance between accuracy and dimension of the basis. The  $GW$ scheme requires the calculation of products  in real space of  KS orbitals with vectors of the polarisability basis. These are represented in  \GWL\ through dedicated additional basis sets of reduced dimensions.

\GWL\ supports only the $\Gamma-$point sampling of the BZ and considers only real wave-functions. However, ordinary $\kk$-point sampling of the BZ can be used for the long-range part of the (symmetrized) dielectric matrix. These terms are calculated by  the \exec{head.x} code. In this way reliable calculations for extended  materials can be performed using quite small simulation cells (with cell edges of the order of 20 Bohr).  Self-consistency is implemented in \GWL, although limited to the quasi-particle energies; the so-called vertex term, arising in the diagrammatic expansion of the self-energy, is not yet implemented.

Usually ordinary $GW$ calculations for transition elements require the explicit inclusion of semicore orbitals  in the valence manifold, resulting in a significantly higher computational cost. To cope with this issue, an approximate  treatment of semicore orbitals has been introduced in  \GWL\ as described in Ref.~ \onlinecite{Umari:12}. In addition to collinear spin polarization, \GWL\ provides a fully relativistic non collinear implementation relying on the scalar relativistic calculation of the screened Coulomb interactions \cite{Umari:14}.  

The \exec{bse.x} code of the \GWL\ package performs BSE calculations and permits to evaluate either the entire  frequency-dependent complex dielectric function through the Lanczos algorithm or a discrete set of excited states and their energies through a conjugate gradient minimization. In contrast to ordinary implementations, \exec{bse.x} scales as $N^3$ instead of $N^4$ with respect to the system size $N$ (\eg, the number of atoms) thanks to the use of maximally localized  Wannier functions for representing the valence manifold \cite{Marsili:2017}. The \exec{bse.x} code, apart from reading the screened Coulomb interaction at zero frequency from a \exec{gww.x} calculation, works as a separate code and uses   the  \qe\  environment. Therefore it could be easily be interfaced with other $GW$ codes.

\paragraph{\SternheimerGW}
The \SternheimerGW\ package calculates the frequency-dependent $GW$ self-energy
and the corresponding quasiparticle corrections at arbitrary
$\kk$-points in the BZ. This feature enables accurate
calculations of band structures and effective masses without resorting
to interpolation. The availability of the complete $GW$ self-energy
(as opposed to the quasiparticle shifts) makes it possible to
calculate spectral functions, for example including plasmon satellites 
\cite{clg15}. The spectral function can be directly compared to
angle-resolved photoelectron spectroscopy (ARPES) experiments. 
In \SternheimerGW\ the screened Coulomb interaction $W$ is evaluated for
wave-vectors in the irreducible BZ by exploiting 
crystal symmetries. Calculations of $G$ and $W$ for multiple
frequencies $\omega$ rely on the use of multishift linear system 
solvers that construct solutions for all frequencies from the solution
of a single linear system~\cite{gcl10, lg13}.
This method is closely related to the Lanczos approach. The
convolution in the frequency domain required to obtain the self energy
from $G$ and $W$ can be performed either  on the real axis or the
imaginary axis. Pad\'e functions are employed to perform approximate
analytic continuations from the imaginary to the real frequency axis; 
the standard Godby-Needs plasmon pole model is also available to
compare with literature results. 
The stability and portability of the \SternheimerGW\ package are verified via a 
test-suite and a Buildbot test-farm (see Sec.~\ref{sec:test-suite}). 

\subsection{Other spectroscopies}
\subsubsection{{\QE-GIPAW}: Nuclear magnetic and electron paramagnetic resonance}

The \QE-GIPAW\ package allows for the calculation of various physical parameters
measured in nuclear magnetic resonance (NMR) and electron paramagnetic
resonance (EPR) spectroscopies. These encompass (i) NMR chemical shift
tensors and magnetic susceptibility, (ii) electric field gradient (EFG)
tensors, (iii) EPR g-tensor, and (iv) hyperfine coupling tensor.

In \QE-GIPAW, the NMR and EPR parameters are obtained  from the orbital
linear response to an external uniform magnetic field. The response
depends critically upon the exact shape of the electronic wavefunctions
near the nuclei. Thus, all-electron wavefunctions are reconstructed
from the pseudo-wavefunctions in a gauge- and translationally invariant
way using the gauge-including projector augmented-wave (GIPAW)
method~\cite{Pickard:2001}. The
description of a uniform magnetic field within periodic boundary
conditions is achieved by the long-wavelength limit ($q \ll 1$) of a  sinusoidally
modulated field in real space. In practice, for each $\kk$ point,
we calculate the first order change of the wavefunctions at $\kk+\qq$,
where $\qq$ runs over a star of 6 points. The magnetic susceptibility and
the induced orbital currents are then evaluated by finite differences,
in the limit of small $q$. The induced magnetic field at the nucleus,
which is the central quantity in NMR, is obtained from the induced current
via the Biot-Savart law. In \QE-GIPAW, the NMR orbital chemical
shifts and magnetic susceptibility can be calculated both for insulators~\cite{Varini:2013}
and for metals~\cite{Avezac:2007} (the additional contribution for metals
coming from the spin-polarization of valence electrons, namely the Knight 
shift, can also be computed but it is not yet ready for production at the
time of writing).
Similarly to the NMR chemical shift, the EPR g-tensor is calculated as the
cross product of the induced current with the spin-orbit operator~\cite{Pickard:2002}.

For the quantities defined in zero
magnetic field, namely the EFG, M\"ossbauer and relativistic
hyperfine tensors, the usual PAW reconstruction of the wavefunctions is sufficient and these are
computed as described in Refs.~\cite{Petrilli98,Zwanziger09}.
The hyperfine Fermi contact term, proportional to the spin density evaluated
at the nuclear coordinates, however requires the relaxation of
the core electrons in response to the magnetization of valence electrons.
We implemented the core relaxation
in perturbation theory, according to Ref.~\onlinecite{Bahramy:2007}.
Basically
we  compute the spherically averaged PAW spin density around each atom. Then
we compute the change of the XC potential, $\Delta V_\text{XC}$, on a radial
grid, and compute in perturbation theory the core radial wavefunction,
both for spin up and spin down. This provides an extra contribution to the
Fermi contact, in most cases opposite in sign to and as large as that of valence electrons.

By combining the quadrupole coupling constants derived from EFG tensors and
hyperfine splittings, electron nuclear double resonance (ENDOR)
frequencies can be calculated. Applications 
highlighting all these features of the \QE-GIPAW\ package can be found in
Ref.~\onlinecite{Bardeleben:2014}. These quantities are also needed to
compute NMR shifts in paramagnetic systems, like novel cathode materials
for Li batteries~\cite{Pigliapochi:2017}.  Previously restricted to
norm-conserving pseudopotentials only, all features are now applicable using any
kind of pseudization scheme and to PAW, following the theory
described in~\cite{Yates:2007}. The use of smooth pseudopotentials allows for the
calculation of chemical shifts in systems with several hundreds of
atoms~\cite{Kucukbenli:2012}.

The starting point of all \QE-GIPAW\ calculations is a previous
calculation of KS orbitals via \PWscf. Hence, much like other
linear response routines, the \QE-GIPAW\ code uses many subroutines
of \PWscf\ and of the linear response module. As usually done in linear
response methods, the response of the unoccupied states is calculated
using the completeness relation between occupied and unoccupied
manifolds~\cite{deGironcoli:1995}. As a result, for insulating as well
as metallic systems, the linear response of the system is efficiently
obtained without the need to include virtual orbitals.

As an alternative to linear response method, the theory of orbital
magnetization via Berry curvature~\cite{Xiao:2005,Thonhauser:2005}
can be used to calculate the NMR~\cite{Thonhauser:2009} and EPR
parameters~\cite{Ceresoli:2010}. Specifically, it can be shown that
the variation of the orbital magnetization $M^{orb}$ with respect to spin flip is
directly related to the g-tensor: $g_{\mu\nu} = g_e - \frac{2}{\alpha
  S} \mathbf{ e}_\mu \cdot \mathbf{M}^{\rm orb}(\mathbf{e}_\nu)$,
where $g_e=2.002319$, $\alpha$ is the fine structure constant,
$S$ is the total spin, $\mathbf{e}$ are Cartesian unit vectors,
provided that the spin-orbit interaction is explicitly considered in
the Hamiltonian. This \emph{converse} method of calculating
the g-tensor has been implemented in an older version of \QE-GIPAW. 
It is especially useful in critical cases where linear response is
not appropriate, \eg, systems with quasi-degenerate HOMO-LUMO levels. A
demonstration of this method applied to delocalized conduction band
electrons can be found in Ref.~\citenum{Behrends:2013}.

The converse method will be shortly ported into the current \QE-GIPAW\ and
we will explore the possibility of computing in a converse way the Knight
shift as the response to a small nuclear magnetic dipole. 
The present version of the code allows for parameter-free calculations
of g-tensors, hyperfine splittings, and ENDOR frequencies also for
systems with total spin $S>1/2$.
Such triplet or even higher-spin states give rise to additional
spin-spin interactions, that can be calculated within the magnetic
dipole-dipole interaction approximation. This interaction results in a
fine structure which can be measured in zero magnetic field. This
so-called zero-field splitting is being implemented following
the methodology described in Ref.~\cite{Bodrog:2014}.

\subsubsection{{\XSPECTRA}: L$_{2,3}$ X-ray absorption edges}

The \XSPECTRA\ code~\cite{PhysRevB.80.075102,PhysRevB.79.045118} 
has been extended to the calculation of X-ray
absorption spectra at the L$_{2,3}$-edges~\cite{PhysRevB.87.205105}. 
The \XSPECTRA\ code uses the
self-consistent charge density produced by \PWscf\ and acts as a post
processing tool~\cite{PhysRevB.80.075102,PhysRevB.79.045118,PhysRevB.66.195107}.
The spectra are calculated for the L$_2$ edge, while the L$_3$ edge
is obtained by multiplying by two (single-particle statistical
branching ratio) the L$_2$ edge spectrum and  by shifting it by the
value of the spin-orbit splitting of the $2p_{1/2}$ core levels of
the absorbing atom. The latter can be taken either from a DFT
relativistic all-electron calculation on the isolated atom,
or from experiments.

In practice, the L$_3$ edge is obtained from the L$_2$ with the
\exec{spectra\_correction.x} tool. Such tool contains a table
of experimental $2p$ spin-orbit splittings for all the elements.
In addition to computing L$_3$ edges, \exec{spectra\_correction.x}
allows one to remove states from the spectrum below a certain energy,
and to convolute the calculated spectrum with more elaborate broadenings.
These operations can be applied to any edge.

To evaluate the X-ray absorption spectrum for a system containing
various atoms of the same species but in different chemical
environments, one has to sum the contribution by each atom.  This
could be the case, for example of an organic molecule containing
various C atoms in inequivalent sites. Such individual contributions
can be computed separately by \XSPECTRA, and the tool
\exec{molecularnexafs.x} allows one to perform their weighted sum taking
into account the proper energy reference (initial and final state
effects)~\cite{10.1021/jp312569q,10.1039/c4cp01625d}.
One should in fact notice that the reference for initial state effects
will depend upon the environment (\eg, the vacuum level for gas phase
molecules, or the Fermi level for molecules adsorbed on a metal).

\subsection{Other lattice-dynamical and thermal properties}

\subsubsection{{\thermopw}: Thermal properties from the quasi-harmonic
  approximation} \label{sec:thermopw}

\thermopw\ \cite{thermopw} is a collection of codes aimed at computing
various thermodynamical quantities in the quasi-harmonic
approximation. The key ingredient is the vibrational contribution, 
$F_{ph}$, to the Helmholtz free energy at temperature $T$: 
\begin{equation}
    F_{ph} = k_B T \sum_{\qq,\nu} \ln \left[ 2 \sinh \left(\frac{\hbar 
        \omega_{\qq\nu} }{ 2k_BT}\right) \right],
\end{equation}
where $\omega_{\qq,\nu}$ are phonon frequencies at wave-vector $\qq$,
$k_B$ is the Boltzmann constant. \thermopw\ works by calling \qe\ routines from \PWscf\ and \PHonon,
that perform one of the following tasks: \emph{i)} compute the DFT total energy and possibly the stress for a 
given crystal structure; \emph{ii)} compute for the same system the electronic band structure along a specified path; \emph{iii)} compute for the same system phonon frequencies at specified wave-vectors.
Using such quantities, \thermopw\ can calculate numerically
the derivatives of the free energy with respect to the external
parameters (\eg, different volumes). Several calls to such routines,
with slightly different geometries, are typically needed in a run.
All such tasks can be independently performed on different groups of
processors (called {\em images}).

When the tasks
carried out by different images require approximately the same time,
or when the amount of numerical work needed to accomplish each task is
easy to estimate a priori, it would be possible to statically assign
tasks to images at the beginning of the run so that images do not need
to communicate during the run. However, such conditions are seldom met
in \thermopw\ and therefore it would be impossible to obtain a good
load balancing between images. \thermopw\ takes advantage of an
engine that controls these different tasks in an asynchronous way,
dynamically assigning tasks to the images at run time.

At the core of \thermopw\ there is a module \var{mp\_asyn}, based on
MPI routines, that allows for asynchronous communication between 
different images. One of the images is the ``master'' and assigns
tasks to the other images (the ``slaves'') as soon as they communicate
that they have accomplished the previously assigned task. The master
image also accomplishes some tasks but once in a while, with
negligible overhead, it checks if there is an image available to do
some work; if so, it assigns to it the next task to do. The code stops
when the master recognizes that all the tasks have been done and
communicates this information to the slaves. The routines of this
communication module are quite independent of the \thermopw\ variables
and in principle can be used in conjunction with other codes to
set up complex workflows to be executed in a massively parallel
environment. It is assumed that each processor of each image reads
the same input and that the only information that the image needs to
synchronize with the other images is which tasks to do.
The design of \thermopw\ makes it easily extensible to the
  calculation of new properties in an incremental way.

\subsubsection{{\thermal2}: phonon-phonon interaction and thermal transport}

Phonon-phonon interaction (ph-ph) plays a role in different physical phenomena: 
phonon lifetime (and its inverse, the linewidth),
phonon-driven thermal transport in insulators or semi-metals,
thermal expansion of materials.
Ph-ph is possible because the harmonic Hamiltonian of ionic motion, of which phonons are stationary states, is only approximate. At first order in perturbation theory we have the third derivative of the total energy with respect to three phonon perturbations, which we compute \emph{ab-initio}. This calculation is performed by the \d3q\ code via the $2n+1$ theorem~\cite{paulattoPRB,lazzeriPRB}.
The \d3q\ code is an extension of the old D3 code, which only allowed the calculation of zone-centered phonon lifetimes and of thermal expansion. The current version can compute the three-phonon matrix element of arbitrary wave-vectors
$D^{(3)}(\qq_1, \qq_2, \qq_3) = \partial^3 E/\partial u_{\qq_1} \partial u_{\qq_2} \partial u_{\qq_3}$, where $u$ are the phonon displacement patterns, the momentum conservation rule imposes $\qq_1 +\qq_2 +\qq_3 = 0$.
The current version of the code can treat any kind of crystal geometry, metals and insulators, both local density and gradient-corrected functionals, and multi-projector norm-conserving pseudopotentials. Ultrasoft pseudopotentials, PAW, spin polarization and non-collinear magnetization are not implemented. Higher order derivative of effective charges\cite{deinzerPRB} are not implemented.

The ph-ph matrix elements, computed from linear response, can be transformed, via a generalized Fourier transform, to the real-space three-body force constants which could be computed in a supercell by finite difference derivation: 
\begin{equation}
D^{(3)} (\qq_1, \qq_2, \qq_3 ) = \sum_{\RR',\RR''} e^{-2i\pi (\RR'\cdot \qq_2 + \RR''\cdot \qq_3)} F^{(3)}(\boldsymbol{0},\RR',\RR''),
\end{equation}
where $F^3(\boldsymbol{0},\RR',\RR'')=\partial^3 E/\partial \tau \partial (\tau'+\RR') \partial (\tau''+\RR'')$ is the derivative of the total energy w.r.t. nuclear positions of ions with basis $\tau$, $\tau'$, $\tau''$ from the unit cells identified by direct lattice vectors $\boldsymbol{0}$ (the origin), $\RR'$ and $\RR''$.
The sum over $\RR'$ and $\RR''$ runs, in principle, over all unit cells, however the terms of the sum quickly decay as the size of the triangle $\boldsymbol{0}-\RR'-\RR''$
increases.
The real-space finite-difference calculation, performed by some external softwares\cite{Li:cpm}, has some advantages: it is easier to implement and it can readily include all the capabilities of the self-consistent code; on the other hand it is much more computationally expensive than the linear-response method we use, its cost scaling with the cube of the supercell volume, or the 9\textsuperscript{th} power of the number of side units of an isotropic system. We use the real-space formalism to apply the sum rule corresponding to translational invariance to the matrix elements. This is done with an iterative method that alternatively enforces the sum rule on the first matrix index and restores the invariance on the order of the derivations. We also use the real-space force constants to Fourier-interpolate the ph-ph matrices on a finer grid, assuming that the contribution from triangles $\boldsymbol{0}-\RR'-\RR''$ which we have not computed is zero; it is important in this stage to consider the periodicity of the system.

From many-body theory we get the first-order phonon linewidth\cite{calandra} ($\gamma_\nu$) of mode $\nu$ at $\qq$, which is a sum over all the possible $N_\qq$'s final and initial states ($\qq'$,$\nu'$,$\nu''$) with conservation of energy ($\hbar\omega$) and momentum ($\qq''=-\qq-\qq'$), Bose-Einstein occupations ($n(\qq,\nu) = (\exp(\hbar\omega_{\qq,\nu}/k_B T)-1)^{-1}$) and an amplitude $V^{(3)}$, proportional to the $D^{(3)}$ matrix element but renormalized with phonon energies and ion masses:
\begin{eqnarray}
\gamma_{\qq,\nu} & =& \frac{\pi}{\hbar^2 N_\qq} \sum_{\qq',\nu',\nu''} \left| V^{(3)}(\qq\nu,\qq'\nu',\qq''\nu'')\right| \times \\
& & \mbox{} \left[
 (1+n_{\qq',\nu'}+n_{\qq'',\nu''})\delta(\omega_{\qq,\nu}-\omega_{\qq',\nu'}-\omega_{\qq'',\nu''})
 +2(n_{\qq',\nu'}-n_{\qq'',\nu''})\delta(\omega_{\qq,\nu}+\omega_{\qq',\nu'}-\omega_{\qq'',\nu''})
\right].\nonumber 
\end{eqnarray}
This sum is computed in the \thermal2 suite, which is bundled with \d3q. A similar expression can be written for the phonon scattering probability which appears in the Boltzmann transport equation. In order to properly converge the integral of the Dirac delta function, we express it as finite-width Gaussian function and use an interpolation grid.
This equation can be solved either exactly or in the single mode approximation (SMA) \cite{callaway:pr}. The SMA is a good tool at temperatures comparable to or larger than the Debye temperature, but is known to be inadequate at low temperatures~\cite{Markov:prb,Markov:sub} or in the case of 2D materials~\cite{broido:prb80,fugallo:nl, Cepellotti:nc}.
The exact solution is computed using a variational form, minimized via a preconditioned conjugate gradient algorithm, which is guaranteed to converge, usually in less than 10 iterations~\cite{fugalloPRB}. 

On top of intrinsic ph-ph events, the \thermal2 codes can also treat isotopic disorder and substitution effects and finite transverse dimension using the Casimir formalism.
In addition to using our force constants from DFpT, the code supports importing 3-body force constants computed via finite differences with the \exec{thirdorder.py} code~\cite{Li:cpm}. Parallelization is implemented with both MPI (with great scalability up to thousand of CPUs) and OpenMP (optimal for memory reduction).

\subsubsection{{\EPW}: Electron-phonon coefficients from Wannier interpolation}

The electron-phonon-Wannier (\EPW) package is designed to calculate electron-phonon coupling using an ultra-fine sampling of the BZ by means of Wannier interpolation. \EPW\ employs the relation between the electron-phonon matrix elements in the Bloch representation $g_{mn\nu}(\kk,\qq)$, and in the Wannier representation,  $g_{ij\kappa\alpha}(\RR,\RR^\prime)$ \cite{Giustino2017},
\begin{equation}
  g_{mn}(\kk,\qq) = \sum_{\RR,\RR^\prime} e^{i(\kk\cdot \RR 
  + \qq\cdot \RR^\prime)}
  \sum_{ij\kappa\alpha} U_{mi \kk+\qq} \,g_{ij\kappa\alpha}(\RR,\RR^\prime) 
  \,U^{\dagger}_{jn\kk} \,u_{\kappa\alpha,\qq\nu},
\end{equation}
in order to interpolate from coarse $\kk$-point and $\qq$-point grids  into dense meshes. In the above expression $\kk$ and $\qq$ represent the electron and phonon wave-vector, respectively, the indices $m,n$ and $i,j$ refer to Bloch states and Wannier states, respectively, and $\RR,\RR^\prime$ are direct lattice vectors. The matrices $U_{mi \kk}$ are unitary transformations and the vector $u_{\kappa\alpha,\qq\nu}$ is the displacements of the atom $\kappa$ along the Cartesian direction $\alpha$ for the phonon of wavevector $\qq$ and branch $\nu$. The interpolation is performed with {\it ab initio} accuracy by relying on the localization of maximally-localized Wannier functions~\cite{marz12}. During its execution \EPW\ invokes the \texttt{Wannier90} software~\cite{Mostofi:2014} in library mode in order to determine the matrices $U_{mi \kk}$ on the coarse $\kk$-point grid.

\begin{figure*}
  \includegraphics{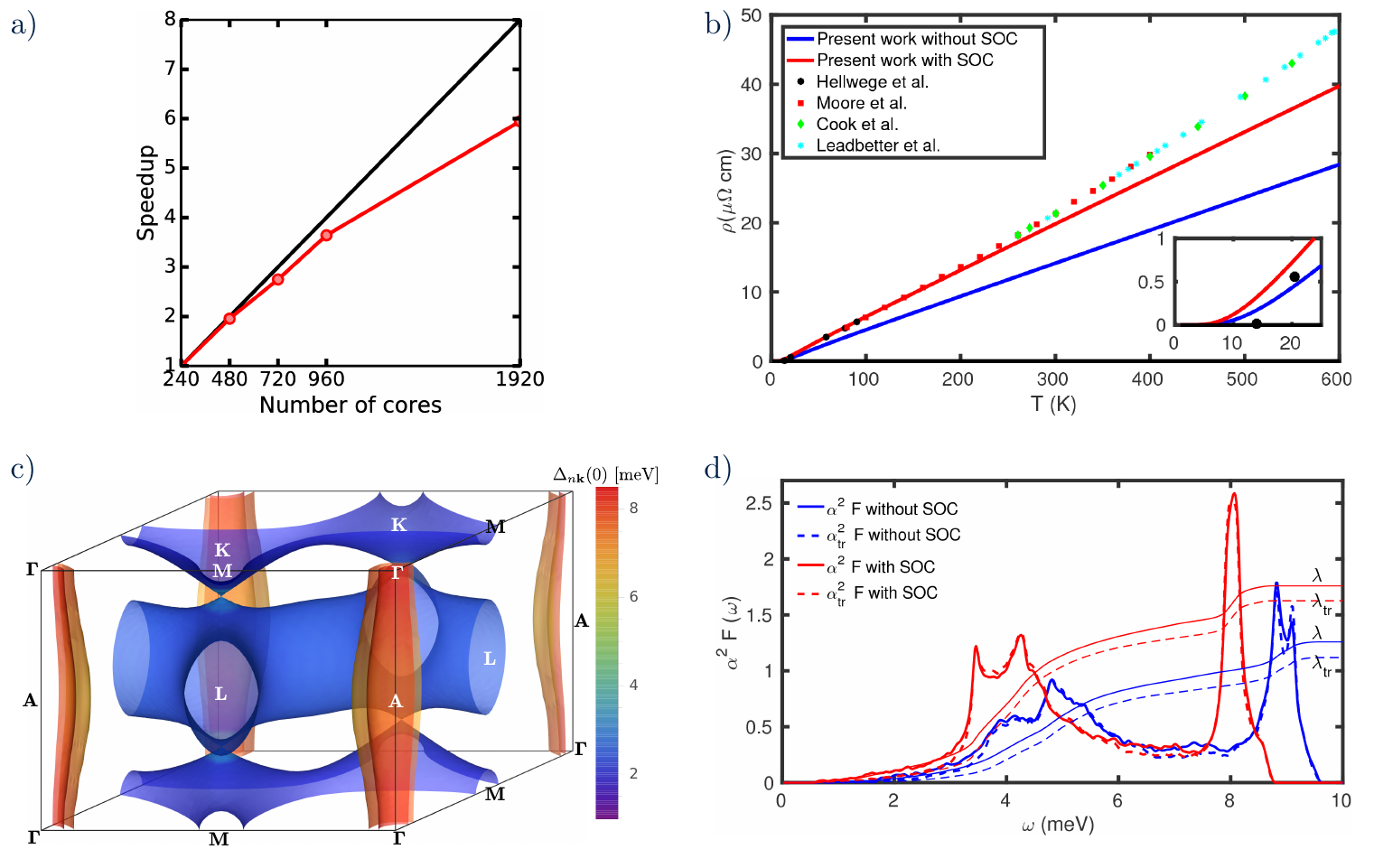}
  \caption{Examples of calculations that can be performed using \EPW. (a) Parallel scaling of \EPW\ on ARCHER Cray XC30. This example corresponds to the calculation of electron-phonon couplings for wurtzite GaN. The parallelization is performed over $\kk$-points using MPI. (b) Calculated temperature-dependent resistivity of Pb by including/neglecting spin-orbit coupling. Reproduced from Ref.~\onlinecite{Ponce:2016}. (c) Calculated superconducting gap function of MgB$_2$, color-coded on the Fermi surface. Reproduced from Ref.~\onlinecite{Ponce:2016}. (d) Eliashberg spectral function $\alpha^2F$ and transport spectral function $\alpha^2F_{\rm tr}$ of Pb. Reproduced from Ref.~\onlinecite{Ponce:2016}.
  }
  \label{fig:epw}
\end{figure*}

\EPW\ can be used to compute the following physical properties: the electron and phonon linewidths arising from electron-phonon interactions; the scattering rates of electrons by phonons; the total, averaged electron-phonon coupling strength; the electrical resistivity of metals, see Fig.~\ref{fig:epw}(b); the critical temperature of electron-phonon superconductors; the anisotropic superconducting gap within the Eliashberg theory,  see Fig.~\ref{fig:epw}(c); the Eliashberg spectral function, transport spectral function, see Fig.~\ref{fig:epw}(d) and the nesting function. The calculation of carrier mobilities using the Boltzmann transport equation in semiconductors is under development.

The \exec{epw.x} code exploits crystal symmetry operations (including time reversal) in order to limit the number of phonon calculations to be performed using \PHonon~to the irreducible wedge of the BZ. The code supports calculations of electron-phonon couplings in the presence of spin-orbit coupling. The current version does not support spin-polarized calculations, ultrasoft pseudopotentials nor the PAW method. As shown in Fig.~\ref{fig:epw}(a), \exec{epw.x} scales reasonably up to 2,000  cores using MPI. A test farm (see Sec.~\ref{sec:test-suite}) was set up to ensure portability of the code on many architecture and 
compilers. Detailed information about the
\EPW\ package can be found in Ref.~\onlinecite{Ponce:2016}. 

\subsubsection{Non-perturbative approaches to vibrational spectroscopies}

Although DFpT is in many ways the state of the art in the simulation of vibrational spectroscopies in extended systems, and in fact one of defining features of \qe, it is sometimes convenient to compute lattice-dynamical properties, the response to macroscopic electric
fields, or combinations thereof (such as \eg, the infrared or Raman activities), using non-perturbative methods. This is so because DFpT
requires the design of dedicated codes, which have to be updated and
maintained separately, and which therefore not always follow the pace
of the implementation of new features, methods, and functionals (such
as \eg, DFT+U, vdW-DF, hybrid functionals, or ACBN0
\cite{Agapito:2015iz}) in their ground-state counterparts. Such a non-perturbative approach is followed in the \FD\ package,   which implements the ``frozen-phonon'' method for the computation of phonons and vibrational spectra: the interatomic Force Constants (IFCs) and electronic dielectric constant are computed as finite differences of forces and polarizations, with respect to finite atomic displacements or external electric fields, respectively \cite{Nardelli_Scientific_Reports_2013,umari2002}. 
IFC's are computed in two steps: first, code \exec{fd.x} generates
the symmetry-independent displacements in an appropriate supercell;
after the calculations for the various displacements are completed,
code \exec{fd\_ifc.x} reads the forces and generates IFC's.
These are further processed in \exec{matdyn.x}, where
non-analytical long-ranged dipolar terms are subtracted out from the IFCs
following the recipe of Ref. \citenum{wang2012}.
The calculation of dielectric tensor and of the Born effective charges
proceeds from the evaluation of the electronic susceptibility
following the method proposed by Umari and Pasquarello
\cite{umari2002}, where the introduction of a non local energy functional 
$E^{\mathbfcal{E}}_{tot}[{\psi}]=E^0[{\psi}]- \mathbfcal{E}\cdot
(\mathbf{P}^{ion}+\mathbf{P}^{el}[{\psi}])$
allows to compute the electronic structure for periodic systems under finite homogeneous electric fields.  $E^0$ is the ground state total energy in the absence of external electric fields; $\mathbf{P}^{ion}$ is the usual ionic polarization, and $\mathbf{P}^{el}$ is given as a Berry phase of the manifold of the occupied bands~\cite{vanderbilt1993}.  The high-frequency dielectric tensor $\epsilon^{\infty}$ is then computed as $\epsilon^{\infty}_{ij}=\delta_{i,j}+4\pi\chi_{ij}$, while Born effective-charge tensors $Z^{*}_{I,ij}$ are obtained as the polarization induced along the direction $i$ by a unit displacement of the $I$-th atom in the $j$ direction; alternatively, as the force induced on atom $I$ by an applied electric field, $\mathbfcal{E}$.

The calculation of the Raman spectra proceeds along similar lines. Within the finite-field approach, the Raman tensor is evaluated in terms of finite differences of atomic forces in the presence of two electric fields \cite{umari2003}. In practice, the tensor $\chi^{(1)}_{ijIk}$ is obtained from a set of calculations combining finite electric fields along different Cartesian directions. $\chi^{(1)}_{ijIk}$ is then symmetrized to recover the full symmetry of the structure under study.

\subsection{Multi-scale modeling}
\subsubsection{{\Environ}: Self-Consistent Continuum Solvation embedding model}
\label{sec:SCCS}

Continuum models are among the most popular multiscale approaches to
treat solvation effects in the quantum-chemistry community
\cite{Tomasi:2005}. In this class of models, the degrees of
freedom of solvent molecules are effectively integrated out and their
statistically-averaged effects on the solute are mimicked by those of
a continuous medium surrounding a cavity in which the solute is
thought to dwell. 
The most important interaction usually handled with
continuum models is the electrostatic one, in which the solvent is
described as a dielectric continuum characterized by its experimental
dielectric permittivity. 

Following the original work of Fattebert and Gygi
\cite{Fattebert:2002} , a new class of continuum models was
designed, in which a smooth transition from the QM-solute region to
the continuum-environment region of space is introduced and defined in
terms of the electronic density of the solute. The corresponding free
energy functional is optimized using a fully variational approach,
leading to a generalized Poisson equation that is solved via a
multi-grid solver\cite{Fattebert:2002}. This approach,
ideally suited for plane-wave basis sets and tailored for MD
simulations, has been featured in the
\qe\ distribution since v. 4.1. This approach
was recently revised\cite{Andreussi:2012}, by defining an
optimally smooth QM/continuum transition, reformulated in terms of
iterative solvers\cite{Fisicaro:2015} and extended to handle
in a compact and effective way non-electrostatic interactions
\cite{Andreussi:2012}. The resulting self-consistent continuum
solvation (SCCS) model, based on a very limited number of
physically justified parameters, allows one to reproduce experimental
solvation energies for aqueous solutions of neutral
\cite{Andreussi:2012} and charged\cite{Dupont:2013}
species with accuracies comparable to or higher than state-of-the-art
quantum-chemistry packages. 

The SCCS model involves different embedding terms, each representing a
specific interaction with an external continuum environment and
contributing to the total energy, KS potential, and interatomic forces
of the embedded QM system. Every contribution may depend explicitly on
the ionic (rigid  schemes) and/or electronic (self-consistent or soft
schemes) degrees of freedom of the embedded system. All the different
terms are collected in the stand-alone
\Environ\ module~\cite{Environ_0.2}. The present discussion refers to
release 0.2 of \Environ, which is compatible with \qe\ starting from
versions 5.1. The module requires a separate input file with the
specifications of the environment interactions to be included and of
the numerical parameters required to compute their effects. 
Fully parameterized and tuned SCCS environments, \eg, corresponding to
water solutions for neutral and charged species, are directly
available to the users. Otherwise individual embedding terms can
be switched on and tuned to the specific physical conditions of the
required environment. Namely, the following terms are currently
featured in \Environ:   
\begin{itemize}
\item Smooth continuum dielectric, with the associated generalized Poisson problem solved via a direct iterative approach or through a preconditioned conjugate gradient algorithm~\cite{Fisicaro:2015}.  
\item Electronic enthalpy functional, introducing an energy term proportional to the quantum-volume of the system and able to describe finite systems under the effect of an applied external pressure\cite{Cococcioni:2005}. 
\item Electronic cavitation functional, introducing an energy term proportional to the quantum-surface able to describe free energies of cavitation and other surface-related interaction terms\cite{Scherlis:2006}.  
\item Parabolic corrections for periodic boundary conditions in aperiodic and partially periodic (slab) systems \cite{Dabo:2008,Andreussi:2014}. 
\item Fixed dielectric regions, allowing for the modelling of complex inhomogenous dielectric environments. 
\item Fixed Gaussian-smoothed distributions of charges, allowing for a simplified modelling of countercharge distributions, \eg, in electrochemical setups.  
\end{itemize}
Different packages of the \qe\ distribution have been interfaced with
the \Environ\ module, including \PWscf, \CP, \PWneb, and \turboTDDFT,
the latter featuring a linear-response implementation
of the SCCS model (see Sec.{\ref{sec:TDDFPT}}).  Moreover, continuum
environment effects are fully compatible with the main features of
\qe, and in particular, with reciprocal space integration and smearing
for metallic systems, with both norm-conserving and ultrasoft
pseudopotentials and PAW, with all XC functionals.  

\subsubsection{QM-MM}
\label{sec:QMMM}

QM-MM was implemented in v.5.0.2 using the method documented in
Ref.~\onlinecite{Ma:2015}. Such methodology accounts for both
mechanical and electrostatic coupling between the QM
(quantum-mechanical) and MM (molecular-mechanics) regions, but not for
bonding interactions (\ie, bonds between the QM and MM regions). In
practice, we need to run two different codes, \qe\ for the QM region
and a classical force-field code for the MM region, that communicate
atomic positions, forces, electrostatic potentials.  

LAMMPS \cite{Plimpton:1995} is the software chosen to deal with the
classical (MM) degrees of freedom. This is a well-known and
well-maintained package, released under an open-source license that  
allows redistribution together with \qe. The communications between
the QM and MM regions use a ``shared memory'' approach: the MM code
runs on a master node, communicates directly via the memory with the
QM code, which is typically running on a massively parallel machine.
Such approach has some advantages: the MM part is typically much
faster than the QM one and can be run in serial execution, wasting no
time on the HPC machine; there is a clear and neat separation between
the two codes, and very small code changes in either codes are
needed. It has however also a few drawbacks, namely: the serial
computation of the MM part may become a bottleneck if the MM region
contains many atoms; direct access to memory is often restricted for
security reasons on HPC machines.   

An alternative approach has been implemented in v.5.4. A single
(parallel) executable runs both the MM and the QM codes. The two codes
exchange data and communicate via MPI. This approach is less elegant
than the previous one and requires a little bit more coding, but its
implementation is quite straightforward thanks also to the changes in
the logic of parallelization mentioned in Sec.~\ref{sec:Overall}. The
coupling of the two codes has required some modifications also to the
\texttt{qmmm} library inside LAMMPS and to the related fix
\texttt{qmmm} (a ``fix'' in LAMMPS is any operation that is applied to
the system during the MD run). In particular,  electrostatic coupling
has been introduced, following the approach described in
Ref.~\onlinecite{Laio:2002}. The great advantage of this approach is
that its performance on HPC machines is as good as the separate
performances of the QM and MM codes. Since LAMMPS is very well
parallelized, this is a significant advantage if the MM region
contains many atoms. Moreover, it can be run without restrictions on
any parallel machine. This new QM-MM implementation is an integral
part of the \qe\ distribution and will soon be included into LAMMPS as
well (the ``fix'' is currently under testing) and it is
straightforward to compile and execute it. 

\subsection{Miscellaneous feature enhancements and additions}

\subsubsection{Fully relativistic projector augmented-wave method}

By applying the PAW formalism to the equations of
relativistic spin density functional theory \cite{rsdft1,rsdft2}, it is
possible to obtain the fully relativistic PAW equations for four-component
spinor pseudo-wavefunctions~\cite{Dalcorso:2010}. In this formalism the
pseudo-wavefunctions can be written in terms of large
$|\tilde \Psi_{i,\sigma}^A\rangle$ and small
$|\tilde \Psi_{i,\sigma}^B\rangle$ components, both two-component
  spinors (the index $\sigma$ runs over the two spin components).
  The latter is of order ${v\over c}$ of the former, where $v$ is of
  the order of the velocity of the electron and $c$ is the speed of
  light.
These equations can be simplified introducing
errors of the order of the transferability error of the pseudopotential
or of order $1/c^2$, depending on
which is the largest. In the final equations only the large components of
the pseudo-wavefunctions appear. The non
relativistic kinetic energy ${\bf p}^2/2m$ ($m$ is the electron mass)
acts on the large component of the pseudo-wavefunctions $|\tilde
\Psi_{i,\sigma}^A\rangle$ in the mesh defined by the FFT grid and the same kinetic
energy is used to calculate the expectation values of the Hamiltonian
between partial pseudo-waves $|\Phi^{I,PS,A}_{n,\sigma} \rangle$.
The Dirac kinetic energy is used instead to calculate the expectation values
of the Hamiltonian between all-electron partial waves
$|\Phi^{I,AE}_{n,\eta} \rangle$ ($\eta$ is a four-component index). In
this manner, relativistic effects are
hidden in the coefficients of the non-local pseudopotential.
The equations are formally very similar to the equations of the
scalar-relativistic case:
\begin{eqnarray}
\sum_{\sigma_2} \Bigg [ \frac{{\bf p}^2 }{ 2m} \delta^{\sigma_1,\sigma_2}  &+& 
\sum_{\eta_1,\eta_2}\int \dr \tilde V_{\rm LOC}^{\eta_1,\eta_2}(\rr) 
\tilde K(\rr)^{\eta_1,\eta_2}_{\sigma_1,\sigma_2} - 
\varepsilon_i S^{\sigma_1,\sigma_2} \nonumber \\ &+& 
\sum_{I,mn} (D_{I,mn}^{1}-\tilde D_{I,mn}^{1}) 
|\beta_{m,\sigma_1}^{I,A} \rangle \langle \beta_{n,\sigma_2}^{I,A} |
\Bigg] | \tilde \Psi_{i,\sigma_2}^A\rangle = 0,
\label{eq:pauli_ks}
\end{eqnarray}
where $D_{I,mn}^{1}$ and $\tilde D_{I,mn}^{1}$ are calculated inside the PAW
spheres:
\begin{eqnarray}
D_{I,mn}^{1} &=& \sum_{\eta_1,\eta_2} 
\langle \Phi^{I,AE}_{m,\eta_1}| T_D^{\eta_1,\eta_2}
+V_{\rm LOC}^{I,\eta_1,\eta_2}| \Phi^{I,AE}_{n,\eta_2} \rangle, \label{d1} \\
\tilde D_{I,mn}^{1} &=& \sum_{\sigma_1,\sigma_2} 
\langle \Phi^{I,PS,A}_{m,\sigma_1}| \frac{{\bf p}^2 }{ 2m}
\delta^{\sigma_1,\sigma_2} +
\tilde V_{\rm LOC}^{I,\sigma_1,\sigma_2} | \Phi^{I,PS,A}_{n,\sigma_2} \rangle 
\nonumber \\
&+&
\sum_{\eta_1,\eta_2}\int_{\Omega_I} \dr \hat Q^I_{mn,\eta_1,\eta_2}(\rr) 
\tilde V_{\rm LOC}^{I,\eta_1,\eta_2}(\rr).
\label{eq:pauli_ks2}
\end{eqnarray}
Here $T_D$ is the Dirac kinetic energy:
\begin{equation}
T_D= c \mbox{\boldmath $\alpha$} \cdot {\bf p} + 
(\mbox{\boldmath $\beta$} -{\bf 1}_{4\times4}) mc^2,
\end{equation}
written in terms of the $4 \times 4$ Hermitian matrices $\mbox{\boldmath $\alpha$}$ and $\mbox{\boldmath $\beta$}$ and $V_{\rm LOC}^{\eta_1,\eta_2}$ is the sum of the local, Hartree, and XC potential ($V_{\rm eff}$) together, in magnetic systems, with the contribution of the XC magnetic field: $V_{\rm LOC}^{\eta_1,\eta_2} (\rr) = V_{\rm eff}(\rr) \delta^{\eta_1,\eta_2} -\mu_B  {\bf B}_{\rm xc} (\rr) \cdot  (\mbox{\boldmath $\beta\Sigma$})^{\eta_1,\eta_2}$. We refer to Ref.~\onlinecite{Dalcorso:2010} for a detailed definition of the partial waves $|\Phi^{I,AE}_{n,\eta}\rangle$, $|\Phi^{I,PS,A}_{n,\sigma}\rangle$ and projectors $|\beta_{m,\sigma}^{I,A} \rangle$, of the augmentation functions $\hat Q^I_{mn,\eta_1,\eta_2}(\rr)$ and $\tilde K(\rr)^{\eta_1,\eta_2}_{\sigma_1,\sigma_2}$, and of the overlap matrix $S^{\sigma_1,\sigma_2}$ and for their rewriting in terms of projector functions that contain only spherical harmonics. Solving these equations it is possible to include spin-orbit coupling effects in electronic structure calculations. In \qe\ these equations are used when input variables \var{noncolin} and \var{lspinorb}  are both \var{.TRUE.} and the PAW data sets are fully relativistic, as those available with the \texttt{pslibrary} project.

\subsubsection{Electronic and structural properties in field-effect configuration}

\begin{figure}
 \includegraphics[width=5cm,clip=]{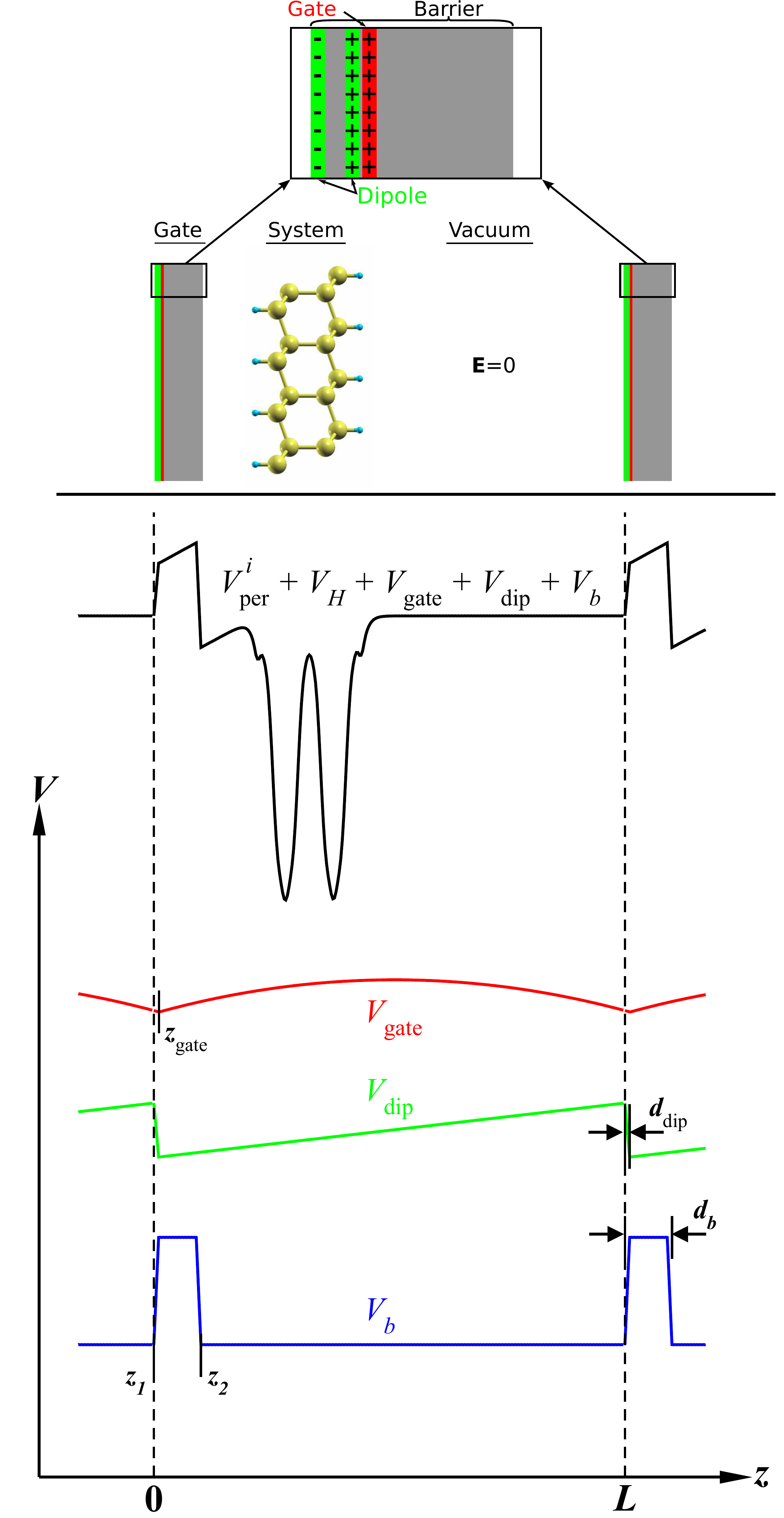}
	\caption{\label{fig:pots} Schematic picture of the planar
          averaged KS potential (without the exchange-correlation
          potential) for periodically repeated, charged slabs.
          The uppermost panel shows a sketch of a gated system.  The different parts of the total KS potential are shown with different color: red -- gate, $V_\mathrm{gate}$, green -- dipole, $V_\mathrm{dip}$, blue -- potential barrier, $V_b$. The position of the gate is indicated by $z_\mathrm{gate}$. The black line shows the sum of $V_\mathrm{gate}$, $V_\mathrm{dip}$, $V_b$, of the ionic potential $V^i_\mathrm{per}$, and of the Hartree potential $V_H$. The length of the unit cell along $\mathbf{\hat{z}}$ is given by $L$.}
\end{figure}

Since \qe\ v.6.0 it is possible to compute the electronic structure
under a field-effect transistor (FET) setup in periodic boundary
conditions \cite{brumme2014}.
In physical FETs, a voltage is applied to a gate electrode, accumulating
charges at the interface between the gate dielectric and a semiconducting system
(see Fig. ~\ref{fig:pots}). The gate electrode is simulated with a
charged plate, henceforth referred to as the {\em gate}. Since the
interaction of this charged plate with its periodic image generates a
spurious nonphysical electric field, a dipolar correction, equivalent
to two planes of opposite charge, is added \cite{bengtsson1999},
canceling out the field on the left side of the gate.
In order to prevent electrons from spilling towards the gate
for large electron doping \cite{topsakal2012}, a potential barrier can be added
to the electrostatic potential, mimicking the effect of the gate
dielectric.

The setup for a system in FET configuration is shown in Fig.~\ref{fig:pots}.
The gate has a charge $n_\mathrm{dop}A$ and the system has
opposite charge. Here $n_\mathrm{dop}$ is the number of doping
electrons per unit area (\ie, negative for hole doping), $A$
is the area of the 
unit cell parallel to the surface. In practice the
gate is represented by an external potential
\begin{eqnarray}
\label{eq:potential}
 V_\mathrm{gate}(\rr) &= 
-2\pi\,n_\mathrm{dop}\,\left(-\left|\overline{\mathrm{z}}\right|+\frac{\overline
{\mathrm{z}}^2}{L}+\frac{L}{6}\right)
\end{eqnarray}
Here $\overline{\mathrm{z}}=z-z_\mathrm{gate}$ with  $\overline{\mathrm{z}}\in\left[-L/2;L/2\right]$ measures the distance from the gate (see Fig.~\ref{fig:pots}). The dipole of the charged system plus the gate is canceled by an electric dipole generated by two planes of opposite charge~\cite{neugebauer1992,bengtsson1999,meyer2001}, placed at $z_\mathrm{dip}-d_\mathrm{dip}/2$ and $z_\mathrm{dip}+d_\mathrm{dip}/2$, in the vacuum region next to the gate ($V_\mathrm{dip}$ in  Fig.~\ref{fig:pots}). Additionally one may include a potential barrier to avoid  charge spilling towards the gate, or as a substitute for the gate dielectric. $V_b(\rr)$ is a periodic function of $z$ defined on the interval $z\in[0,L]$ as equal to a constant $V_b$ for $z_1<z<z_2$ and zero elsewhere. Figure \ref{fig:pots} shows the resulting total potential (black line). The following additional variables are needed: $z_\mathrm{gate}$, $z_1$, $z_2$, and $V_0$. In the code these variables are named \var{zgate},  \var{block\_1}, \var{block\_2}, and \var{block\_height}, respectively. The dipole corrections and the gate are activated by the options  \var{dipfield=.true.} and \var{gate=.true.}. In order to enable the potential barrier and the relaxation of the system towards it, the new input parameters \var{block} and \var{relaxz}, respectively, have to be set to \var{.true.} More details about the implementation can be found in Ref. \citenum{brumme2014}.

\subsubsection{Cold restart of Car-Parrinello molecular dynamics}

In the standard Lagrangian formulation of \emph{ab initio} molecular
dynamics \cite{Car:1985}, the coefficients of KS molecular orbitals
over a given basis set (\ie. their Fourier coefficients, in the
case of plane waves) are treated as classical degrees of freedom
obeying Newton's equations of motion that derive from a suitably
defined extended Lagrangian. This Lagrangian is obtained from the
Born-Oppenheimer total energy by augmenting it with a fictitious
electronic kinetic-energy term and relaxing the constraint that the
molecular orbitals stay at each instant of the trajectory in their
instantaneous KS ground state. The idea is that, by choosing a
suitably small \emph{fictitious electronic mass}, the thermalization
time of the electronic degrees of freedom can be made much longer than
the typical simulation times, so that if the system is prepared in its
electronic KS ground state at the start of the simulation, the
electronic dynamics would follow almost adiabatically the nuclear one
all over the simulation, thus effectively mimicking a \emph{bona fide}
Born-Oppenheimer dynamics.  

While in Car-Parrinello MD both the physical nuclear and fictitious
electronic velocities are determined by the equations of motion on a
par, the question still remains as to how choose them at the start of
the simulation. Initial nuclear velocities are dictated by physical
considerations (\eg, thermal equilibrium) or may be taken from a
previously interrupted MD run. Electronic velocities (\ie,
  the time derivatives of the KS molecular orbitals), instead,
are not available when the simulation is started from scratch.
and are not independent of the physical nuclear ones,
but are determined by the adiabatic time evolution of the system.
Moreover, the
projection over the occupied-state manifold of the electronic
velocities, $\dot \psi_v^\parallel\doteq \hat P \dot\psi_v$ is
ill-defined because the KS ground-state solution is defined modulo a
unitary transformation within this manifold.
This means that the starting electronic velocities may not be
  simply obtained as finite differences of KS orbitals at times
  $t=0$ and $t=\Delta t$. Here and in the following
$\hat P$ indicates the projector over the occupied-state manifold, and
$\hat Q=1-\hat P$ its complement (\ie\ the projector over the
virtual-orbital manifold).

The component of the electronic velocities
over the virtual-state manifold, $\dot \psi_v^\perp\doteq \hat
Q\dot\psi_v$, is instead well defined and can be formally  written
using standard first-order perturbation theory:
\begin{equation}
  \dot\psi^\perp_v(\rr) = \sum_{c}\psi_c(\rr)\frac{\langle\psi_c|\dot V_{KS}|\psi_v\rangle}{\epsilon_v-\epsilon_c}, \label{eq:VelocityPT}
\end{equation}
where $v$ and $c$ indicate occupied (\emph{valence}) and virtual
(\emph{conduction}) states, respectively, $\epsilon_n$ the
corresponding orbital energies, and $\dot V_{KS}$ is the time
derivative of the KS potential, $V_{KS}$. $\dot V_{KS}$ is the linear
response of $V_{KS}$ to the perturbation in the external potential
determined by an infinitesimal displacement of the nuclei along a MD
trajectory: $\dot V_\mathrm{ext}(\rr)=\sum_\RR \frac{\partial
  v_\RR(\rr-\RR)}{\partial\RR}\cdot\dot\RR$, where $ v_\RR(\rr-\RR) $
is the bare ionic pseudopotential of the atom at position $\RR$ and
$\dot\RR$ its velocity. Electronic velocities can conveniently be
initialized to the values given by Eq. \eqref{eq:VelocityPT}, which
are those that minimize their norm and, hence, the initial
\emph{electronic temperature}, which is defined as the sum of the
squared norms of the electronic velocities.

While this could in
principle be done using density-functional perturbation theory
\cite{Baroni:1987,Baroni:2001}, it is more
convenient to compute them numerically, following the procedure
described below. 
At $t=0$ the KS molecular orbitals are initialized from a ground-state
computation, performed with whatever method is available or preferred
(standard SCF calculation or global optimization, such as \eg, with
conjugate gradients \cite{Stich:1989}). The KS molecular orbitals that
would result from a perfectly adiabatic propagation at $t=\Delta t$
are then determined from a second ground-state computation, performed
after half a ``velocity-Verlet" MD step, \ie, at nuclear positions
$\RR(\Delta t)=\RR(0)+\dot\RR(0)\Delta t$. The initial velocities are
then obtained from the relation: 
\begin{equation}
  \dot\psi_v^\perp=\hat{\dot P}\psi_v, \label{eq:VelocityN}
\end{equation}
which is obtained by simply differentiating
the definition of occupied-state projector, $\hat P\psi_v=\psi_v$. The
right-hand side of Eq. \eqref{eq:VelocityN} is finally easily computed
by subtracting from each KS orbital at time $t=0$, its component over the
occupied-state manifold at $t=\Delta t$ and dividing by $\Delta t$. 

\subsubsection{Optimized tetrahedron method}

The integration over $\kk$-points in the BZ is a crucial step
in the calculation of the electronic structure of a periodic system,
affecting not only the ground state but linear response as well.
This is especially true for metallic systems where the integrand is
discontinuous at the Fermi level. Even more problematic is the
integration of Dirac delta functions, such as those appearing in the
density of states (DOS), partial DOS and in the electron-phonon coupling
constant.

\qe\ has always implemented a variety of ``smearing'' methods, in which
the delta function is replaced by a function of finite width (\eg, a
Gaussian function, or more sophisticated choices).
It has also always implemented the linear tetrahedron method 
\cite{Jepson19711763} with the correction proposed by Bl\"ochl
\cite{PhysRevB.49.16223}, in which the BZ is divided into
tetrahedra and the integration is performed analytically by linear
interpolation of KS eigenvalues in each tetrahedron. Such method is however
limited in its convenience and range of applicability: in fact the linear
interpolation systematically overestimates convex functions, thus making
the convergence against the number of $\kk$-points slow. The linear tetrahedron
method was thus mostly restricted to the calculation of DOS and partial DOS.

Since \qe\ v.6.1, the optimized tetrahedron method \cite{PhysRevB.89.094515}
is implemented. Such method overcomes the drawback of the linear tetrahedron
method using an interpolation that accounts for the curvature of the
interpolated function. The optimized tetrahedron method has better
convergence properties and an extended range of applicability: in addition
to the calculation of the ground-state charge density, DOS and partial DOS,
it can be used in linear-response calculation of phonons and
of the electron-phonon coupling constant.

\subsubsection{Wyckoff positions}

In \qe\ the crystal geometry is traditionally specified by a Bravais lattice index (called \var{ibrav}), by the crystal parameters (\var{celldm}, or \var{a}, \var{b}, \var{c}, \var{cosab}, \var{cosac}, \var{cosbc}) describing the unit cell, and by the positions of {\em all} atoms in the unit cell, in crystal or Cartesian axis.

Since v.5.1.1, it is possible to specify the crystal geometry in
crystallographic style\cite{zadra}, according to the notations of the
International Tables of Crystallography (ITA)\cite{ITA:2005}.  A
complete description of the crystal structure is obtained by
specifying the space-group number according to the ITA and the
positions of {\em symmetry-inequivalent} atoms only in the unit
cell. The latter can be provided either in the crystal axis of the
conventional cell, or as {\em Wyckoff positions}: a set of special
positions, listed in the ITA for each space group, that can be fully
specified by a number of parameters, none to three depending upon the
site symmetry. Table 1 reports a few examples of accepted syntax.
\begin{table}[h]
   \caption{
     Examples of valid syntax for Wyckoff positions. $C$ is the
     element name, followed by the Wyckoff label of the site (number
     of equivalent atoms followed by a letter identifying the site),
     followed by the site-dependent parameters needed to fully specify
     the atomic positions.}
    {ATOMIC\_POSITIONS sg}\\
    \begin{tabular}{l l l l l}
    $C$ & 1a\\
    $C$ & 8g&   x \\
    $C$ & 24m&  x& y \\
    $C$ & 48n&  x& y& z\\
    $C$ & &     x& y& z\\
    \end{tabular}
\end{table}
The code generates the symmetry
operations for the specified space group and applies them to
inequivalent atoms, thus finding all atoms in the unit cell. 

For some crystal systems there are alternate descriptions in the ITA,
so additional input parameters may be needed to select the desired
one. For the monoclinic system the ``c-unique'' orientation is
the default and \var{bunique=.TRUE.} must be specified in input
if the ``b-unique'' orientation is desired. For some space groups
there are two possible choices of the origin. The  origin appearing first  in the ITA 
is chosen by default, unless \var{origin\_choice=2} is
specified in input. Finally, for trigonal space groups the atomic
coordinates can be referred to the rhombohedral or to the hexagonal
Bravais lattices. The default is the rhombohedral lattice, so
\var{rhombohedral=.FALSE.} must be specified in input to use the
hexagonal lattice. 

A final comment for centered Bravais lattices: in the crystallographic
literature, the {\em conventional} unit cell is usually assumed.
\qe\ however assumes the {\em primitive} unit cell, having a smaller
volume and a smaller number of atoms, and discards atoms outside the
primitive cell. Auxiliary code \exec{supercell.x}, available in
\thermopw\ (see Sec.\ref{sec:thermopw}), prints all atoms in the
conventional cell when necessary.

\section{Parallelization, modularization, interoperability and stability}

\subsection{New parallelization levels}
\label{sec:newpara}

The basic modules of \qe\ are characterized by a hierarchy of
parallelization levels, described in Ref.\onlinecite{QE:2009}.
Processors are divided into groups, labeled by
a MPI communicator. Each group of processors distributes a specific
subset of computations. The growing diffusion of HPC machines based on
nodes with many cores (32 and more) makes however pure MPI
parallelization not always ideal: running one MPI  process per core
has a high overhead, limiting performances. It is often convenient to
use mixed MPI-OpenMP parallelization, in which a small number of MPI
processes per node use OpenMP threads, either explicitly (\ie, with
compiler directives) or implicitly (\ie, via calls to OpenMP-aware
library). Explicit OpenMP parallelization, originally confined to
computationally intensive FFT's, has been
extended to many more parts of the code. 

One of the challenges presented by massively parallel machine is to
get rid of both memory and CPU time bottlenecks, caused respectively
by arrays that are not distributed across processors and by
non-parallelized sections of code.
It is especially important to distribute all arrays and parallelize
all computations whose size/complexity increases
with the dimensions of the unit cell or of the basis
set. Non-parallelized computations hamper ``weak'' scalability,
that is, parallel performance while increasing both the system size
and the amount of computational resources, 
while non-distributed arrays may become an unavoidable RAM bottleneck
with increasing problem size. ``Strong'' scalability
(that is, at fixed problem size and increasing number of CPUs) is
even more elusive than weak 
scalability in electronic-structure calculations, requiring, in
addition to systematic distribution of computations, to keep to the
minimum the ratio between time spent in communications and in
computation, and to have a nearly perfect load balancing. In order to
achieve strong scalability, the key is to add more parallelization
levels and to use algorithms that permit to overlap communications and
computations. 

For what concerns memory, notable offenders are arrays of scalar
products between KS states $\psi_i$:  $O_{ij} = \langle\psi_i|\widehat
O|\psi_j\rangle$, where $\widehat O$ can be either the Hamiltonian or
an overlap matrix; and scalar products between KS states and
pseudopotential projectors $\beta$, $p_{ij} =
\langle\psi_i|\beta_j\rangle$. The size 
of such arrays  grows as the square of the size of the cell. Almost
all of them are now distributed across processors of the
``linear-algebra group'', that is, the group of processors taking care
of linear-algebra operations on matrices. The most expensive of such
operations are subspace diagonalization (used in \PWscf\ in the
iterative diagonalization) and iterative orthonormalization (used by
\CP). In both cases, a parallel dense-matrix diagonalization on
distributed matrix is needed. In addition to ScaLAPACK, \qe\ can now
take advantage of newer ELPA libraries (Ref.~\onlinecite{ELPA:2014}),
leading to significant performance improvements. 

The array containing the plane-wave representation, $c_{\kk,n}(\GG)$,
of KS orbitals is typically the largest array, or one of the largest.
While plane waves are already 
distributed across processors of the ``plane-wave group" as defined in
Ref.~\onlinecite{QE:2009}, it is now possible to distribute KS
orbitals as well. Such a parallelization level is located between the
$\kk$-point and the plane-wave parallelization levels. The
corresponding MPI communicator defines a subgroup of the ``$\kk$-point
group" of processors and is called ``band  group communicator". In the
\CP\ package, band parallelization is implemented  for almost all
available calculations. Its usefulness is better appreciated in
simulations of large cells --- several hundreds of atoms and more ---
where the number of processors required by memory distribution
would be too large to get good scalability from plane-wave
parallelization only. 

In \PWscf, band parallelization is implemented for calculations using
hybrid functionals. The standard algorithm to compute Hartree-Fock
exchange in a plane-wave basis set (see Sec.~\ref{sec:exx})
contains a double loop on bands
that is by far the heaviest part of computation. A first form of
parallelization, described in Ref.~\onlinecite{Varini:2013}, was
implemented in v.5.0. In the latest version, this has been
superseded by parallelization of pairs of bands, Ref.~\onlinecite{Nersc:2017}.
Such algorithm is compatible with the
``task-group" parallelization level (that is: over KS states in the
calculation of $V\psi_i$ products) described in Ref.~\onlinecite{QE:2009}. 

In addition to the above-mentioned groups, that are globally defined
and in principle usable in all routines, there are a few additional
parallelization levels that are local to specific routines. Their goal
is to reduce the amount of non-parallel computations that may become
significant for many-atom systems. An example
is the calculation of DFT+U (Sec.~\ref{sec:dft+U}) terms in energy and
forces, Eqs.~(\ref{eq:Uener}) and (\ref{eq:Uforce}) respectively.
In all these expressions, the calculation of the scalar products
between valence and atomic wave functions is in principle the most
expensive step: for $N_b$ bands and $N_{pw}$ plane waves,
${\cal O}(N_{pw}N_b)$ floating-point operations are required
(typically, $N_{pw} \gg N_b$). The calculation  of these terms is however
easily and effectively parallelized, using standard  matrix-matrix
multiplication routines and summing over MPI processes with a
\var{mpi\_reduce} 
operation on the plane-wave group. The sum over $\kk$-points can be
parallelized on the $\kk$-point group. The remaining sums over band
indices $\nu$ and Hubbard orbitals $I,m$ may however 
require a significant amount of non-parallelized computation if the
number of atoms  with a Hubbard $U$ term is not small. The sum over
band indices is thus parallelized by simply distributing
bands over the plane-wave group. This is a convenient choice because
all processors of the plane-wave group are available once the
scalar  products are calculated. The addition of band parallelization
speeds up the computation of such terms by a significant factor.
This is  especially important for  Car-Parrinello dynamics,
requiring the calculation of forces at each time step, when a sizable
number of Hubbard manifolds is present. 

\subsection{Aspects of interoperability}
\label{sec:Interoperability}

One of the original goals of \qe\ was to assemble different pieces
of rather similar software into an integrated software suite.
The choice was made to focus on the following four aspects:
input data formats, output data files, installation mechanism,
and a common base of code. While work on the first three aspects
is basically completed, it is still ongoing on the fourth. 
It was however realized that a different form of integration ---
interoperability, \ie, the possibility to run \qe\ along with other
software --- was more useful to the community of users than tight 
integration. There are several reasons for this, all rooted in new
or recent trends in computational materials science. We mention in
particular the usefulness of interoperability for
\begin{enumerate}
\item excited-states calculations using many-body perturbation theory,
  at various levels of sophistication: $GW$, TDDFT, BSE (\eg,
  \yambo{}~\cite{Marini:2009}, \sax{}~\cite{MartinSamos:2009}, or
  \BerkeleyGW{}~\cite{Deslippe:2012});
\item calculations using quantum Monte Carlo methods;
\item configuration-space sampling, using such algorithms as nudged 
   elastic band (NEB), genetic/evolutionary algorithms, meta-dynamics;
\item inclusion of quantum effects on nuclei via path-integral Monte Carlo;
\item multi-scale simulations, requiring different theoretical approaches,
   each valid in a given range of time and length scale, to be used 
   together;
 \item high-throughput, or ``exhaustive'', calculations
   (\eg, \AiiDA~\cite{Pizzi:2016,Mounet:2016} and AFLOW$\pi$~\cite{AFLOWPI})
   requiring automated  submission, analysis, retrieval of a large
   number of jobs;
\item ``steering'', \ie, controlling the computation in real time using
   either a graphical user interface (GUI) or an interface in a high-level
   interpreted language (\eg, python).
\end{enumerate}
It is in principle possible, and done in some cases, to implement all
of the above into \qe, but this is not always the best practice.
A better option is to use \qe\ in conjunction with external software
performing other tasks. 

Cases 1 and 2 mentioned above typically use as starting step the
self-consistent solution of KS equations, so that what is needed is the
possibility for external software to read data files produced by the main
\qe\ codes, notably the self-consistent code \PWscf\ and the molecular
dynamics code \CP.

Cases 3 and 4 typically require many self-consistent calculations at
different atomic configurations, so that what is needed is the possibility
to use the main \qe\ codes as ``computational engine'', \ie, to call
\PWscf\ or \CP\ from an external software, using atomic configurations
supplied by the calling code.

The paradigmatic case 5 is QM-MM (Sec.\ref{sec:QMMM}), requiring an
exchange of data, notably: atomic positions, forces, and some
information on the electrostatic potential, between \qe\ and the MM
code -- typically a classical MD code. 

Case 6 requires easy access to output data from one simulation, and
easy on-the-fly generation of input data files as well. This is also
needed for case 7, which however may also require a finer-grained
control over computations performed by \qe\ routines: in the most
sophisticated scenario, the GUI or python interface should be able to
perform specific operations ``on the fly'', not just running an entire
self-consistent calculation. This scenario relies upon the existence
of a set of application programming interfaces (API's) for calls to
basic computational tasks. 

\subsection{Input/Output and data file format}

On modern machines, characterized by fast CPU's and large RAM's, disk input/output (I/O) may become a bottleneck and should be kept to a strict minimum. Since v.5.3 both \PWscf\ and \CP\ do not perform by default any I/O at run time, except for the ordinary text output (printout), for checkpointing if required or needed, and for saving data at the end of the run. The same is being gradually extended to all codes. In the following, we discuss the case of the final data writing. 

The original organization of output data files (or more exactly, of the output data directory) was based on a formatted ``head'' file, with a XML-like syntax, containing general information on the run, and on binary data files containing the KS orbitals and the charge density. We consider the basic idea of such approach still valid, but some improvements were needed. On one hand, the original head file format had a number of small issues---inconsistencies, missing pieces of relevant information---and used a non-standard syntax, lacking a XML ``schema'' for validation. On the other hand, data files suffered from the lack of portability of Fortran binary files and had to be transformed into text files, sometimes very large ones, in order to become usable on a different machine. 

\subsubsection{XML files with schema}

Since v.6.0, the ``head'' file is a true XML file using a consistent
syntax, described by a XML schema, that can be easily parsed with
standard XML tools. It also contains complete information on the run,
including all data needed to reproduce the results, and on the correct
execution and exit status. This aspect is very useful for
high-throughput applications, for databasing of results and for
verification and validation purposes.

The XML file contains an input section and can thus be used as input
file, alternative to the still existing text-based input. It
supersedes the previous XML-based input, introduced several years ago,
that had a non-standard syntax, different from and incompatible with
the one of the original head file. Implementing a different input is
made easy by the clear separation existing between the reading and
initialization phases: input data is read, stored in a separate
module, copied to internal variables. 

The current XML file can be easily parsed and generated using standard
XML tools and is especially valuable in conjunction with GUI's.
The schema can be found at the URL:\\
\texttt{http://www.quantum-espresso.org/ns/qes/qes-1.0.xsd}.

\subsubsection{Large-record data file format}

Although not as I/O-bound as other kinds of calculations, electronic-structure simulations may produce a sizable amount of data, either intermediate or  needed for further processing. The largest array typically contains the plane-wave representation of KS orbitals; other sizable arrays contain the charge and spin density, either in reciprocal or in real space. In parallel execution using MPI, large arrays are distributed across processors, so one has two possibilities: let each MPI process write its own slice of the data (``distributed'' I/O), or collect the entire array on a single processor before writing it (``collected'' I/O). In distributed I/O, coding is straightforward and efficient, minimizing file size and achieving some sort of I/O parallelization. A global file system, accessible to all MPI processes, is needed. The data is spread into many files that are directly usable only by a code using exactly the same distribution of arrays, that is, exactly the same kind of parallelization. In collected I/O, the coding is less straightforward. In order to ensure portability, some reference ordering, independent upon the number of processors and the details of the parallelization, must be provided. For large simulations, memory usage and communication pattern must be carefully optimized when a distributed array is collected into a large array on a single processor.

In the original I/O format, KS orbitals were saved in reciprocal space, in either distributed or collected format. For the latter, a reproducible ordering of plane waves (including the ordering within shells of plane waves with the same module), independent upon parallelization details and machine-independent, ensures data portability. Charge and spin density were instead saved in real space and in collected format. In the new I/O scheme, available since v.6.0, the output directory is simplified, containing only the XML data file, one file per $\kk$-point with KS orbitals, one file for the charge and spin density. Both files are in collected format and both quantities are stored in reciprocal space. In addition to Fortran binary, it is possible to write data files in HDF5 format\cite{hdf5}. HDF5 offers the possibility to write structured record and portability across architectures, without significant loss in performances; it has an excellent support and is the standard for I/O in other fields of scientific computing. Distributed I/O is kept only for checkpointing or as a last-resort alternative. 

In spite of its advantages, such a solution has still a bottleneck in
large-scale computations on massively parallel machines: a single
processor must read and write large files. Only in the case of
parallelization over $\kk$-points is I/O parallelized in a straightforward
way. More general solutions to implement parallel I/O using parallel
extensions of HDF5 are currently under examination in view of enabling
\qe\ towards ``exascale'' computing (that is: towards ${\cal
  O}(10^{18})$ floating-point operations per second).

\subsection{Organization of the distribution}
\label{sec:Overall}

Codes contained in \qe\ have evolved from a small set of original
codes, born with rather restricted goals, into a much larger
distribution via continuous additions and extensions. Such a process -
presumably common to most if not all scientific software projects -
can easily lead to uncoordinated growth and to bad decisions that
negatively affect maintainability. 

\subsubsection{Package re-organization and modularization}

In order to make the distribution easier to maintain, extend and
debug, the distribution has been split into 
\begin{itemize}
 \item[a.] base distribution, containing common libraries, tools
   and utilities, core packages \PWscf, \CP, \PostProc, plus some commonly
		used additional packages, currently: \atomic, \texttt{PWgui},
		\PWneb, \PHonon, \XSPECTRA,
   \turboTDDFT, \turboEELS, \GWL, \EPW;
 \item[b.] external packages such as \sax{}~\cite{MartinSamos:2009}, \yambo{}~\cite{Marini:2009}, \Wannier90{}~\cite{Mostofi:2014}, \WanT{}~\cite{Calzolari:2004,Ferretti:2007},
   that are automatically downloaded and installed on demand.
\end{itemize}
The directory structure now explicitly reflects the structure of
\qe\ as a ``federation'' of packages rather than a monolithic one:
a common base distribution plus additional packages,
each of which fully contained into a subdirectory. 

In the reorganization process, the implementation of the NEB algorithm
was completely rewritten, following the paradigm sketched in
Sec.~\ref{sec:Interoperability}. \PWneb\ is now a separate package
that implements the NEB algorithm, using \PWscf\ as the computational
engine. The separation between the NEB algorithm and the
self-consistency algorithm is quite complete: \PWneb\ could be adapted
to work in conjunction with a different computational engine with a
minor effort. 

The implementation of meta-dynamics has also been re-considered.
Given the existence of a very sophisticated and
well-maintained package\cite{Bonomi:2009} \Plumed\ for all kinds of
meta-dynamics 
calculations, the \PWscf\ and \CP\ packages have been adapted to work
in conjunction with \Plumed{} v.1.x, removing the old internal
meta-dynamics code. In order to activate meta-dynamics,  a patching
process is needed, in which a few specific ``hook'' routines are
modified so that they call routines from \Plumed{}.

\subsubsection{Modular parallelism}

The logic of parallelism has also evolved towards a more modular
approach. It is now possible to have all \qe\ routines working
inside a MPI communicator, passed as argument to an initialization
routine. This allows in particular the calling code to have its own
parallelization level, invisible to \qe\ routines; the latter can thus
perform independent calculations, to be subsequently processed by the
calling code. For instance: the ``image'' parallelization level, used
by NEB calculations, is now entirely managed by \PWneb\ and no longer in
the called \PWscf\ routines. Such a feature is very useful for coupling
external codes to \qe\ routines. To this end, a general-purpose
library for calling \PWscf\ or \CP\ from external codes (either
Fortran or C/C++ using the Fortran 2003 ISO C binding standard) is
provided in the directory \texttt{COUPLE/}. 

\subsubsection{Reorganization of linear-response codes}
\label{sec:codereorg} 

All linear-response codes described in Secs.~\ref{sec:lrexc} and
\ref{sec:ACFD} share as basic computational step the self-consistent
solution of linear systems $Ax=b$ for different perturbations $b$,
where the operator $A$ is derived from the KS Hamiltonian $H$ and the
linear-response potential.
Both the perturbations and the methods of solution differ
by subtle details, leading to a plethora of routines, customized to
solve slightly different versions of the same problem. Ideally, one
should be able to solve any linear-response problem by using a
suitable library of existing code. To this end, a major restructuring
of linear-response codes has been started. Several routines have been
unified, generalized and extended. They have been collected into the
same subdirectory, \texttt{LR\_Modules}, that will be the container of
``generic'' linear-response routines. Linear-response-related packages
now contain only code that is specific to a given perturbation or property
calculation. 

\subsection{\qe\ and scripting languages}

A desirable feature of electronic-structure codes is the ability to be
called from a high-level interpreted scripting language. Among the
various alternatives, python has emerged in the last years due to its
simple and powerful syntax and to the availability
of numerical extensions (NumPy). Since v.6.0, an interface
between \PWscf\ and the path integral MD driver
i-PI~\cite{Ceriotti:2014} is available and distributed together with
\qe. Various implementations of an interface between \qe\ codes and
the atomic simulation environment (ASE) \cite{Bahn:2002} are also
available. In the following we briefly highlight the integration of 
\qe\ with \AiiDA, the \pwtk\ toolkit for \PWscf, and 
the \QEmodes\ package for user-friendly editing of \qe\ with the \texttt{Emacs} editor
\cite{emacs}.

\subsubsection{\AiiDA: a python materials' informatics infrastructure}

\begin{figure}
  \centering
  \includegraphics[width=0.8\textwidth]{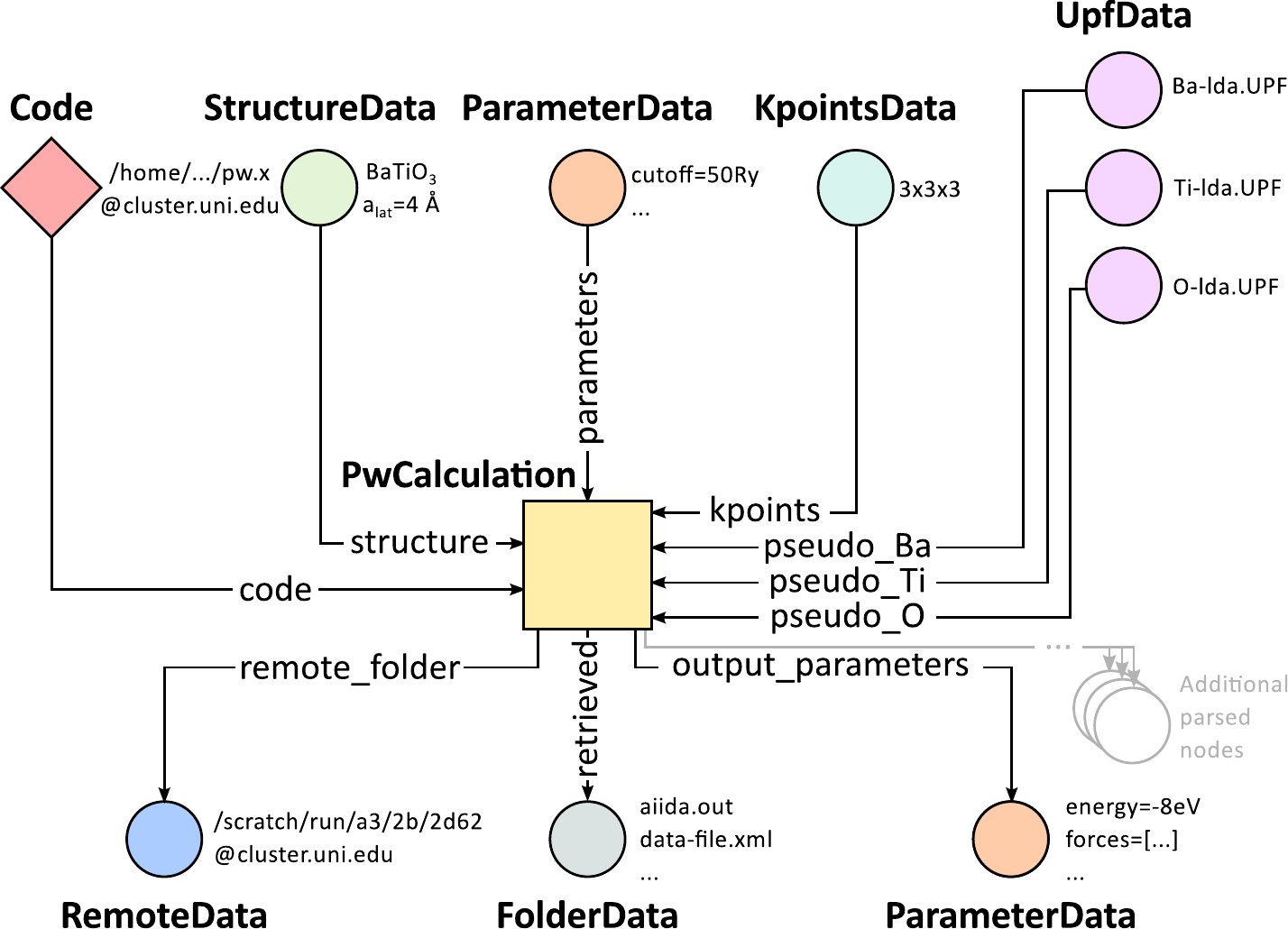}
  \caption{A simple \AiiDA\ directed acyclic graph for a \qe\ calculation using \PWscf\   
  (square), with all the input nodes (data, circles; code executable, diamond) and all the output nodes that the daemon creates and connects automatically.}
  \label{fig:aiida_graph}
\end{figure}

\AiiDA\ \cite{Pizzi:2016} is a comprehensive python infrastructure 
aimed at accelerating, simplifying, and organizing major efforts in computational 
science, and in particular computational materials science, with a close integration 
with the \qe\ distribution. \AiiDA\ is structured around the four pillars of the 
ADES model (Automation, Data, Environment, and Sharing, Ref. \citenum{Pizzi:2016})), 
and provides a practical and efficient implementation of all four. In particular, 
it aims at relieving the work of a computational scientist from the tedious and 
error-prone tasks of running, overseeing, and storing hundreds or more of calculations 
daily (Automation pillar), while ensuring that strict protocols are in place to 
store these calculations in an appropriately structured database that preserves 
the provenance of all computational steps (Data pillar). This way, the effort 
of a computational scientist can become focused on developing, curating, or 
exploiting complex workflows (Environment pillar) that calculate in a robust 
manner e.g. the desired materials properties of a given input structure, 
recording expertise in reproducible sequences that can be progressively 
perfected, while being able to share freely both the workflows and the 
data generated with public or private common repositories (Sharing). 
\AiiDA\ is built using an agnostic structure that allows to interface 
it with any given code --- through plugins and a plugin repository --- or 
with different queuing systems, transports to remote HPC resources, and 
property calculators. In addition, it allows to use arbitrary object-relational 
mappers (ORMs) as backends (currently, Django and SQLAlchemy are supported). 
These ORMs map the \AiiDA\ objects (``Codes'', ``Calculations'' and ``Data'') 
onto python classes, and lead to the representation of calculations through 
Directed Acyclic Graphs (DAGs) connecting all objects with directional arrows; 
this ensures both provenance and reproducibility of a calculation. As an example, 
in Fig.~\ref{fig:aiida_graph} we present a simple DAG representing a
  \PWscf\ calculation on BaTiO$_3$.

\subsubsection{\Pwtk: a toolkit for \PWscf}

The \pwtk, standing for \texttt{{\bf PW}scf {\bf T}ool{\bf K}it}, 
is a Tcl scripting interface for \PWscf\ set of programs contained in the
  \qe\ distribution. It aims at providing a flexible and
  productive framework.  The basic philosophy of \pwtk\ is to lower
the learning curve by using syntax that closely
matches the input syntax of \qe.
\Pwtk\ features include:
(i) assignment of default values of input variables on a
project basis, (ii) reassignment of input
variables on the fly, (iii) stacking of input data, (iv) math-parser,
(v) extensible and hierarchical configuration (global, project-based,
local), (vi) data retrieval functions (\ie, either loading the data
from pre-existing input files or retrieving the data from output
files), and (vii) a few predefined higher-level tasks, that
consist of several seamlessly integrated calculations.
\Pwtk\ allows to easily automate large number of
  calculations and to glue together different computational tasks, where
  output data of preceding calculations serve as input for subsequent
  calculations.
\Pwtk\ and related documentation can be downloaded from
\texttt{http://pwtk.quantum-espresso.org}.

\subsubsection{\QEmodes}

\begin{figure*}[ht]
  \begin{tabular}[t]{cc}
    (a) \includegraphics[height=0.45\textwidth]{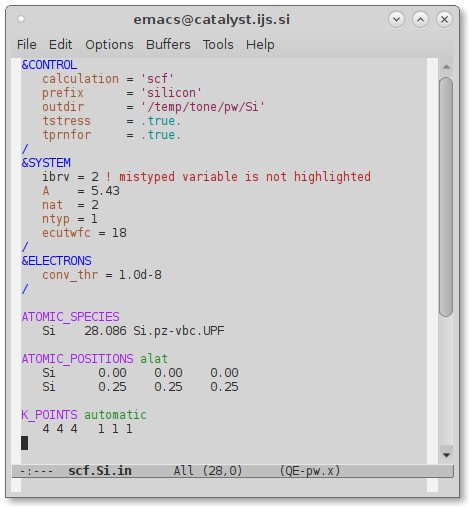}
    & (b) \includegraphics[height=0.45\textwidth]{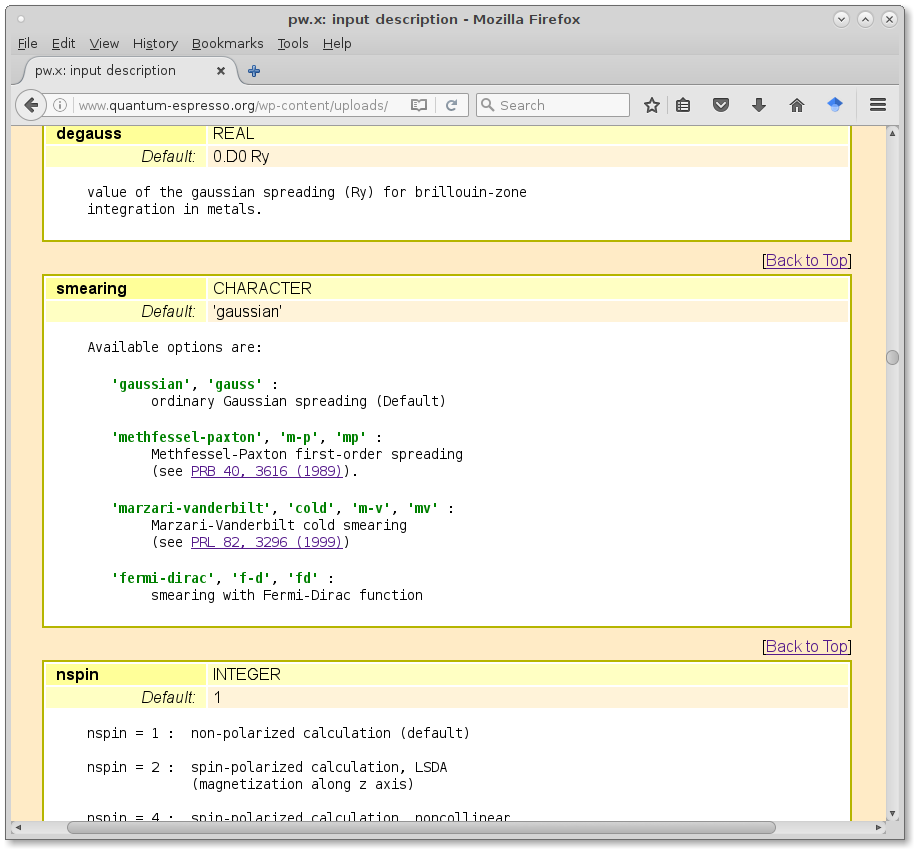}\\
  \end{tabular}
  \caption{(a) \texttt{pw.x} input file opened in \texttt{Emacs} with
    \pwmode\ highlighting the following elements: namelists and their
    variables (blue and brown), cards and their options (purple and
    green), comments (red), string and logical variable values
    (burgundy and cyan, respectively). A mistyped variable (\ie,
    \var{ibrv} instead of \var{ibrav}) is not highlighted. (b) An
    excerpt from the \texttt{INPUT\_PW.html} file, which describes the
    \exec{pw.x} input file syntax. Both the \QEmodes\ and the
    \texttt{INPUT\_PW.html} are automatically generated from the \qe's
    internal definition of the input file syntax.}
  \label{fig:qe-modes}
\end{figure*}

The \QEmodes\ package is an open-source collection of \texttt{Emacs}
major-modes for making the editing of \qe\ input files
easier and more comfortable with
\texttt{Emacs}. The package provides syntax highlighting (see
Fig.~\ref{fig:qe-modes}a), auto-indentation, auto-completion, and
a few utility commands, such as \texttt{M-x}~{\it
  prog}\texttt{-insert-template} that inserts a respective input file
template for the \texttt{\it prog} program (\eg, \texttt{pw},
\texttt{neb}, \texttt{pp}, \texttt{projwfc}, \texttt{dos},
\texttt{bands}).
The \QEmodes\ are
aware of all namelists, variables, cards, and options that are explicitly documented in the \texttt{INPUT\_}{\it
  PROG}\texttt{.html} files, which describe the respective input
file syntax (see Fig.~\ref{fig:qe-modes}b), where {\it PROG}
stands for the uppercase name of a given program of \qe. The
reason for this is that both \texttt{INPUT\_}{\it PROG}\texttt{.html}
files and \QEmodes\ are automatically generated by the internal
\texttt{helpdoc} utility of \qe.
    
\subsection{Continuous Integration and testing} \label{sec:test-suite}

The modularization of \qe~reduces the extent of code duplication, thus improving
code maintainability, but it also creates interdependencies 
between the modules so that changes to one part of the code may impact other 
parts. In order to monitor and mitigate these side effects we developed a test-suite for
non-regression testing. Its purpose is to increase code stability by 
identifying and correcting those changes that break established functionalities.
The test-suite relies on a modified version of python script \texttt{testcode}
\cite{testcode}.

\begin{figure}
  \centering
  \includegraphics{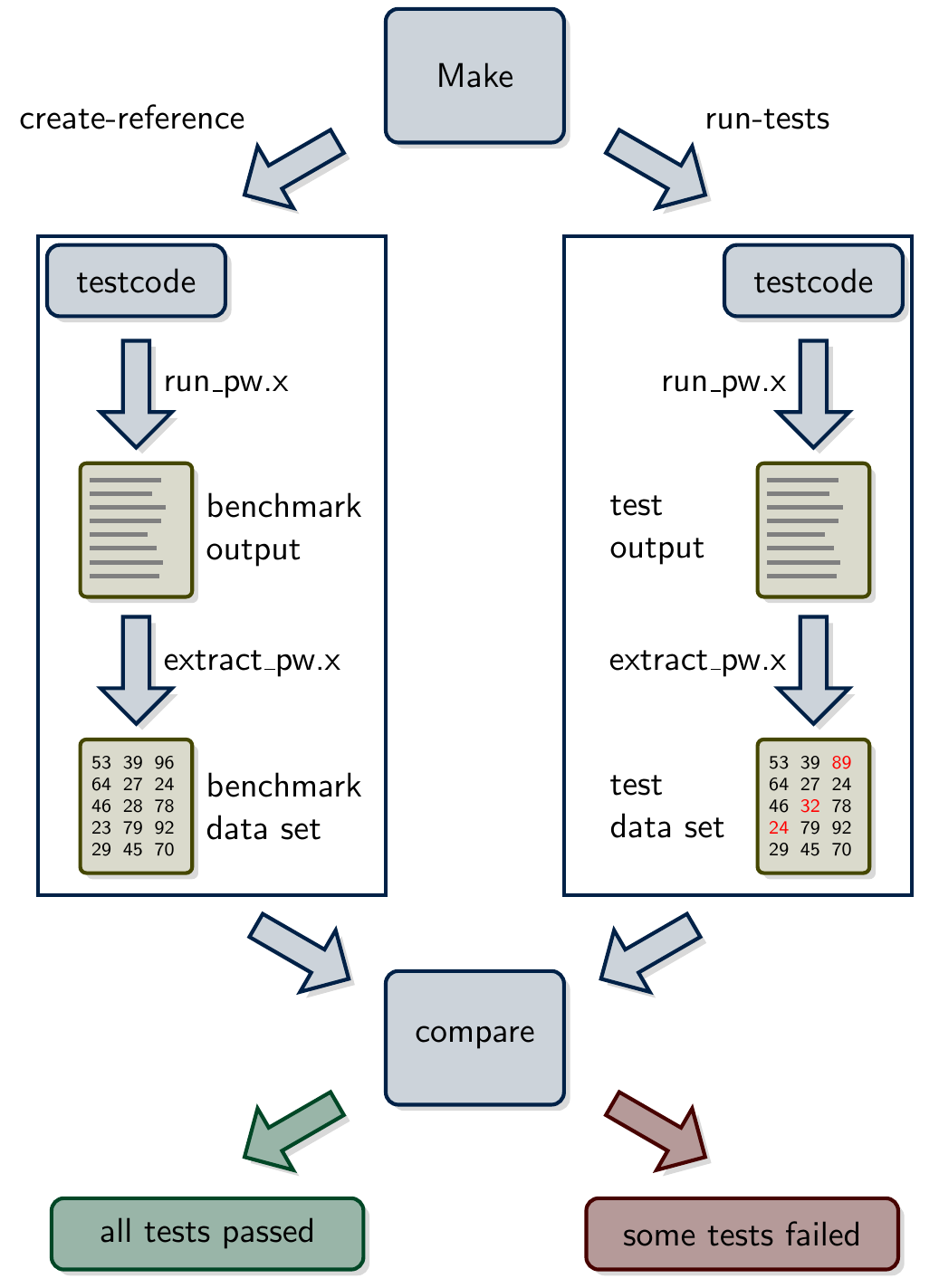}
  \caption{
  (Color online) Layout of the \qe~test-suite. The program \texttt{testcode} runs 
  \qe~executables, extracts numerical values from the output files, and compares
  the results with reference data. If the difference between these data exceeds
  a specified threshold, \texttt{testcode} issues an error indicating that
  a recent commit might have introduced a bug in parts of the code.}
  \label{fig:test-suite}
\end{figure}
The layout of the test-suite is illustrated in Fig.~\ref{fig:test-suite}.
The suite is invoked via a Makefile that accepts several options
to run sequential or parallel tests or to test one particular
feature of the code. The test-suite runs the various executables of \qe,
extracts the numerical data of interest, compares them to reference data,
and decides whether the test is successful using specified thresholds.
At the moment, the test-suite contains 181 tests for \texttt{PW},
14 for \texttt{PH}, 17 for \texttt{CP}, 43 for \texttt{EPW}, and 6
for \texttt{TDDFpT} covering 43\%, 30\%, 29\%, 63\% and 25\% of the
blocks, respectively. Moreover, 60\%, 44\%, 47\%, 76\% and 32\% of the
subroutines in each of these codes are tested, respectively.

The test-suite also contains the logic to automatically create reference
data by running the relevant executables and storing the output in a benchmark
file. These benchmarks are updated only when new tests are added or bugfixes
modify the previous behavior.

The test-suite enables automatic testing of the code using several
\texttt{Buildbot} test farms. The test farms monitor the code repository
continuously, and trigger daily builds in the night after every new commit.
Several compilers (Intel, GFortran, PGI) are tested both in serial and in 
parallel (openmpi, mpich, Intel mpi and mvapich2) execution with different
mathematical libraries (LAPACK, BLAS, ScaLAPACK, FFTW3, MKL, OpenBlas). 
More information can be found at \url{test-farm.quantum-espresso.org}.

 The official mirror of the development version of \qe~(\texttt{https://github.com/QEF/q-e}) employs
a subset of the test-suite to run \texttt{Travis CI}. This tool rapidly identifies erroneous
commits and can be used to assist code review during a pull request.

\section{Outlook and conclusions}

This paper describes the core methodological developments
and extensions of \qe\ that have become available, or are
about to be released, after Ref. \citenum{QE:2009}
appeared. The main goal of \qe\ to provide an efficient and
extensible framework to perform simulations with well-established 
approaches and to develop new methods remains firm, and it has nurtured
an ever growing community of developers and contributors.

Achieving such goal, however, becomes increasingly challenging. On
one hand, computational methods become ever more complex and
sophisticated, making it harder not only to implement them on a
computer but also to verify the correctness of the implementation
(for a much needed initial effort on verification of electronic-structure codes
based on DFT, see Ref. \onlinecite{Lejaeghereaad3000}).
On the other hand, exploiting the current technological innovations in computer
hardware can requires massive changes to software and even algorithms. This is
especially true for the case of ``accelerated'' architectures (GPUs and
the like), whose exceptional performance can translate to actual calculations
only after heavy restructuring and optimization. The complexity of 
existing codes makes a rewrite for new architectures a challenging choice, 
and a risky one given the fast evolution of computer architectures.

We think that the main directions followed until now in the
development of \qe\ are still valid, not only for new methodologies,
but also for adapting to new computer architectures and future ``exascale'' 
machines. Namely, we will continue pushing towards code reusability, by
removing duplicated code and/or replacing it with routines performing
well-defined tasks, by identifying the time-intensive sections of the code 
for machine-dependent optimization, by having documented APIs with a predictable
behavior and with limited dependency upon global variables, and we will
continue to optimize performance and reliability.
Finally, we will push towards extended interoperability
with other software, also in view of its usefulness for data exchange
and for cross-verification, or to satisfy the needs of high-throughput calculations.

Still, the investment in the development and maintenance of state-of-the-art scientific software 
has historically lagged behind compared to the investment in the applications that use such software, and one wonders
is this the correct or even forward-looking approach given the strategic importance of such tools, their impact, their
powerful contribution to open science, and their full and complete availability to the entire community.
In all of this, the future of materials simulations appear ever more bright\cite{Marzari:2016}, and the
usefulness and relevance of such tools to accelerating invention and discovery in science and
technology is reflected in its massive uptake by the community at large.

\textbf{Acknowledgments}.
This work has been partially funded by the European Union through the
\textsc{MaX} Centre of Excellence (Grant No. 676598) and by the {\sc Quantum
ESPRESSO} Foundation. 
SdG acknowledges support from the EU Centre of Excellence \texttt{E-CAM}
(Grant No. 676531). OA, MC, NC, NM, NLN, and IT acknowledge support from the SNSF National
Centre of Competence in Research MARVEL, and from the PASC Platform for
Advanced Scientific Computing. TT acknowledges support from NSF Grant No.\ DMR-1145968.
SP, MS, and FG are supported by the Leverhulme Trust (Grant RL-2012-001).
MBN acknowledges support by DOD-ONR (N00014-13-1-0635, N00014-11-1-0136, 
N00014-15-1-2863) and the Texas Advanced Computing Center at the University 
of Texas, Austin.
RD acknowledges partial support from Cornell University through start-up funding and the Cornell Center for Materials Research (CCMR) with funding from the NSF MRSEC program (DMR-1120296). This research used resources of the Argonne Leadership Computing Facility at Argonne National Laboratory, which is supported by the Office of Science of the U.S. Department of Energy under Contract No. DE-AC02-06CH11357. This research also used resources of the National Energy Research Scientific Computing Center, which is supported by the Office of Science of the U.S. Department of Energy under Contract No. DE-AC02-05CH11231.

\bibliography{QEbiblio}
\bibliographystyle{apsrev4-1}

\end{document}